\documentclass[
preprint,
amsmath,
amssymb,
aps,
pra,
showkeys,
12pt
]{revtex4-1}

\usepackage{graphicx}
\usepackage{newtxtext}
\usepackage{newtxmath}
\usepackage{natbib}
\usepackage{hyperref}
\usepackage{subfigure}
\usepackage{overpic}
\usepackage{physics}
\usepackage{siunitx}
\usepackage[cymk]{xcolor}
\hypersetup{
    colorlinks = true,
    urlcolor   = blue,
    citecolor  = black,
}
\usepackage[mathlines]{lineno}

\newcommand{\RomanNumeralCaps}[1]

\newcommand{\bs}{\boldsymbol}

\newcommand{\mc}{\mathcal}
\newcommand{\mr}{\mathrm}
\newcommand{\mbb}{\mathbb}

\newcommand{\EQ}{\begin{equation}}
\newcommand{\EN}{\end{equation}}
\newcommand{\EQA}{\begin{eqnarray}}
\newcommand{\ENA}{\end{eqnarray}}

\newcommand{\lrr}[1]{\left(#1\right)}
\newcommand{\lrs}[1]{\left[#1\right]}

\newcommand{\lrN}[1]{\left\Vert#1\right\Vert}

\newcommand\Rey{\mbox{\textit{Re}}}  
\newcommand\etal{\mbox{\textit{et al.}}}

\begin{document}

\title{Generative prediction of flow fields around an obstacle using the diffusion model}

\author{Jiajun Hu} 
\affiliation{State Key Laboratory for Turbulence and Complex Systems, College of Engineering, Peking University, Beijing 100871, China}
\author{Zhen Lu} 
\email{zhen.lu@pku.edu.cn}
\affiliation{State Key Laboratory for Turbulence and Complex Systems, College of Engineering, Peking University, Beijing 100871, China}
\author{Yue Yang} 
\email{yyg@pku.edu.cn}
\affiliation{State Key Laboratory for Turbulence and Complex Systems, College of Engineering, Peking University, Beijing 100871, China}
\affiliation{HEDPS-CAPT, Peking University, Beijing 100871, China}

\date{\today}

\begin{abstract}

We propose a geometry-to-flow diffusion model that utilizes obstacle shape as input to predict a flow field around an obstacle.
The model is based on a learnable Markov transition kernel to recover the data distribution from the Gaussian distribution. 
The Markov process is conditioned on the obstacle geometry, estimating the noise to be removed at each step, implemented via a U-Net. 
A cross-attention mechanism incorporates the geometry as a prompt. 
We train the geometry-to-flow diffusion model using a dataset of flows around simple obstacles, including circles, ellipses, rectangles, and triangles. 
For comparison, two CNN-based models and a VAE model are trained on the same dataset.
Tests are carried out on flows around obstacles with simple and complex geometries, representing interpolation and generalization on the geometry condition, respectively. 
To evaluate performance under demanding conditions, the test set incorporates scenarios including crosses and the characters `PKU.' 
Generated flow fields show that the geometry-to-flow diffusion model is superior to the CNN-based models and the VAE model in predicting instantaneous flow fields and handling complex geometries. 
Quantitative analysis of the accuracy and divergence demonstrates the model's robustness.

\end{abstract}


\maketitle


\section{Introduction}\label{sec:intro}

Understanding the interaction between fluid and solid objects is crucial for optimizing a broad range of engineering applications~\cite{Shelley2011}. 
Flow over different obstacles, for which geometry information is typically available, is common and important.
Computational fluid dynamics (CFD) serves as a key tool for revealing the complex fluid-solid interactions~\cite{Tong2021, Tong2022} and supporting efficient design processes. 
However, high-fidelity CFD requires expertise to create accurate computational meshes, especially for complex geometries~\cite{Perot2011}, and they demand considerable computing resources~\cite{Kern2024}.
These challenges have motivated the exploration of CFD-accelerating approaches.

Machine learning (ML) has become a powerful tool for accelerating CFD~\cite{Vinuesa2022, Lino2023, Brunton2024}.
Efforts include reconstructing flow fields~\cite{Fukami2019, Kim2021, Fukami2023}, performing interpolations~\cite{Kochkov2021, Hu2024}, and generating inflow conditions~\cite{Yousif2022, Yousif2023}.
Following these developments, predicting the flow fields has gained increasing attention.
Convolutional neural network (CNN)~\cite{Hasegawa2020, Nakamura2021, Shirzadi2022, Hada2024}, graph neural networks~\cite{BelbutePeres2020, Lino2022, Xie2024gnn, Gao2024gnn}, and physics-informed neural networks (PINNs)~\cite{Karniadakis2021, Jin2021, Yang2021} have been explored to predict flow fields around obstacles. 
Recent researches underscore the importance of generalization and out-of-distribution validation in neural partial differential equation solvers, including Fourier neural operators~\cite{Li2021, Wen2022, Wang2024}, deep operator network~\cite{Lu2021, Jin2023, Bai2024}, and PDEformer~\cite{Ye2024}.
Nevertheless, ensuring robust performance in out-of-distribution scenarios remains a significant challenge~\cite{Fukami2024}, especially when dealing with diverse obstacle geometries, necessitating effective methods that can handle complex shapes. 

Generative models, known for their strong generalization capability~\cite{Biever2023}, have also been applied to flow field prediction~\cite{Lee2019,Gundersen2021,Wang2021,Wang2023,Yang2023,Kim2024,SoleraRico2024,Abaidi2024}. 
By learning underlying data distributions, these models can generate new samples for CFD tasks. 
Early works employed generative adversarial networks (GANs) for unsteady cylinder flow~\cite{Lee2019} and variational autoencoders (VAEs) for flow around cylinders~\cite {Gundersen2021} and airfoils~\cite{Wang2021}. 
Other studies integrated VAEs with transformers to capture chaotic flow patterns~\cite{SoleraRico2024}, or adopted GAN-based frameworks to handle data-scarce scenarios~\cite{Wang2023} and multi-variable flow generation~\cite{Abaidi2024}. 
More recently, diffusion-based methods~\cite{Yang2023}, which have excelled in high-quality image generation~\cite{Dhariwal2021}, show promise for more stable training and complex feature representation, making them particularly appealing for challenging flow configurations.

The diffusion models~\cite{Yang2023} have emerged as powerful probabilistic methods, with two main variants: the denoising diffusion probabilistic model (DDPM)~\cite{SohlDickstein2015, Ho2020} and score-based generative models~\cite{Song2020}. 
Li \etal~\cite{Li2024} applied the diffusion model to generate Lagrangian turbulence, later extending to weighted particle trajectories~\cite{Li2024b} and conditional turbulent signal reconstruction~\cite{Li2024c}. 
Lienen \etal~\cite{Lienen2024} developed a conditional diffusion model for 3D turbulent flow generation past obstacles with different cross-section shapes, and Saydemir \etal~\cite{Saydemir2024} further explored time-varying flows. 
Gao \etal~\cite{Gao2024} incorporated Bayesian conditions into diffusion models to achieve long-time and long-span turbulent flow generation. 
Du \etal~\cite{Du2024} combined a conditioned neural field with a latent diffusion model to efficiently generates high-fidelity spatiotemporal turbulent flows under various conditions. 
Liu \etal~\cite{Liu2024} extended this framework to generate inflow conditions for varying Reynolds numbers.

Despite recent advances in diffusion models for fluid dynamics, a gap exists in their application to 2D flow fields around streamwise-aligned geometries. 
While Lienen \etal~\cite{Lienen2024} made valuable contributions with diffusion models for 3D turbulent flows, their focus on obstacles with cross-sections orthogonal to the flow direction addresses fundamentally different fluid dynamics problems. 
The 2D streamwise-aligned configuration represents a distinct case worthy of specialized attention, as it directly isolates the relation between an obstacle's profile and critical flow phenomena such as separation points and vortex dynamics. 
Current research has primarily limited its scope to airfoil flows~\cite{Ning2025} with restricted geometric parameters, leaving the broader space of arbitrary geometries largely unexplored. 
This highlights the need for a specialized diffusion-based framework for diverse streamwise-aligned profiles. Its progressive denoising process is particularly valuable for handling the complex, non-linear relations between geometry and resulting flow patterns.

We develop a geometry-to-flow (G2F) diffusion model in which the obstacle geometry serves as a prompt to generate the flow field. 
The G2F diffusion model is trained using elementary geometries, and its performance is assessed on both simple and complex shapes to examine interpolation and generalization. 
The results demonstrate its capability to produce flow fields for different obstacle geometries, including those outside the training distribution.
The diffusion model can accelerate CFD workflows by providing high-quality initial field approximations that improve solver convergence efficiency and reduce computational overhead in traditional numerical methods~\cite{Ning2025}.
The remaining manuscript is organized as follows. 
Section 2 introduces the model framework, neural network (NN) structure, and dataset for training and testing. 
Section 3 details the validation results for various geometries. 
The final section provides concluding remarks.

\section{Geometry prompt to flow field output}
\label{sec:model}

\subsection{DDPM}\label{subsec:DDPM}

\begin{figure}[ht]
  \centerline{\includegraphics[width=\textwidth]{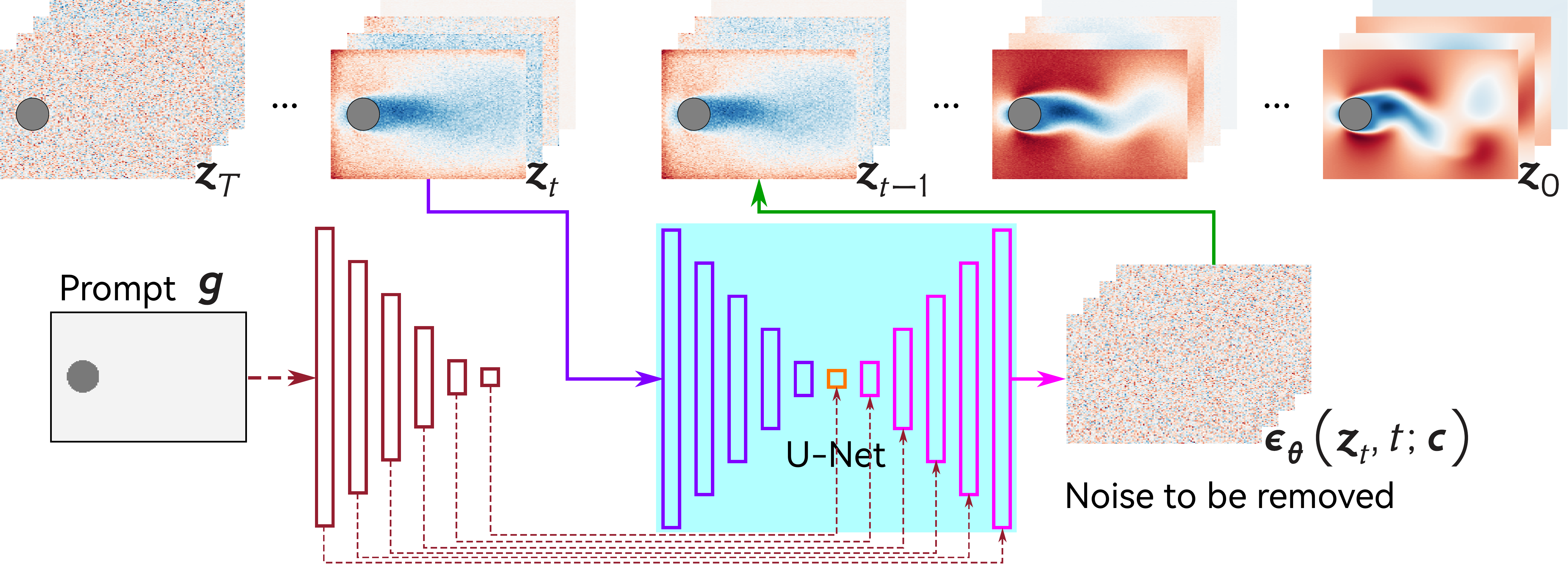}}
  \caption{
  Schematic for the reverse diffusion process in the G2F diffusion model. It generates a sample by removing noises through a U-Net in steps. 
  The geometry prompt is injected into the U-Net via an attention mechanism, which takes the geometry as the prompt to control the generation process.
  }
\label{fig:DDPM}
\end{figure}

We use the obstacle geometry as a prompt to generate a flow around the obstacle as output through the DDPM.  
Figure~\ref{fig:DDPM} provides an overview of our G2F diffusion model. 
There are two diffusion processes in the DDPM -- the forward diffusion perturbs data to noise, and the reverse diffusion converts noise back to data~\cite{SohlDickstein2015}. 
The DDPM learns a probabilistic representation of the underlying data, thereby facilitating the generation of high-quality samples that resemble the target distribution.

For data $\bs{z}_0$ sampled from a distribution $q(\bs{z}_0)$, the forward diffusion utilizes a Markov transition kernel 
\EQ\label{eq:forward}
    q\lrr{\bs{z}_t\vert \bs{z}_{t-1}} = \mc{N}\lrr{\bs{z}_t;\sqrt{1-\beta_t} \bs{z}_{t-1}, \beta_t\bs{I}}, \quad t = 1, \dots, T
\EN
to inject noise, generating a sequence of noisy samples, $\bs{z}_1, \dots, \bs{z}_T$ in total $T$ steps, where $\beta_t \in \lrr{0, 1}$ is the diffusion rate, $\mc{N}$ denotes the Gaussian distribution, and $\bs{I}$ is the identity matrix.
For $T\rightarrow \infty$, $q\lrr{\bs{z}_T} \approx \mc{N}\lrr{\bs{0}, \bs{I}}$, allowing to approximate $\bs{z}_T$ as a Gaussian vector.

To generate samples $\bs{z}_0$, DDPM conducts a reverse diffusion process on the Gaussian vector $\bs{z}_T$ using a learnable Markov transition kernel 
\EQ\label{eq:reverse}
    p_{\bs\theta}\lrr{\bs{z}_{t-1}\vert\bs{z}_t; \bs{c}} = \mc{N}\lrs{\bs{z}_{t-1}; \bs{\mu}_{\bs\theta}\lrr{\bs{z}_t, t}, \bs{\Sigma}_{\bs\theta}\lrr{\bs{z}_t, t}, \bs{c}},
\EN
where $\bs\theta$ denotes model parameters, $\bs{c}$ denotes the condition, the mean $\bs{\mu}_{\bs\theta}$ and variance $\bs{\Sigma}_{\bs\theta}$ are parameterized by the neural network.
The parameters $\bs\theta$ are trained to make the reverse diffusion process a good approximation of the forward one, minimizing the Kullback-Leibler (KL) divergence~\cite{Kullback1951} between $q\lrr{\bs{z}_0, \bs{z}_1, \dots, \bs{z}_T}$ and $p_{\bs\theta}\lrr{\bs{z}_0, \bs{z}_1, \dots, \bs{z}_T; \bs{c}}$.
To optimize training, Ho \etal~\cite{Ho2020} proposed a loss function based on the mean-square-error (MSE) of the predicted noise,
\EQ\label{eq:lossMSE}
    \mc{L} = \mbb{E}_{t,\bs{z}_0,\bs{\epsilon}}
    \lrs{ \lrN{ \bs{\epsilon} - \bs{\epsilon}_{\bs\theta} \lrr{\bs{z}_t, t; \bs{c}} }^2 },
\EN
where $\mbb{E}$ denotes the expectation, and $\bs{\epsilon}_{\bs\theta}\lrr{\bs{z}_t, t; \bs{c}}$ is a NN predicting the noise vector $\bs{\epsilon}\sim \mc{N}\lrr{\bs{0},\bs{I}}$.
Note that a single NN for $\bs{\epsilon}_{\bs\theta}\lrr{\bs{z}_t, t; \bs{c}}$ is trained for all diffusion steps. 

\subsection{Geometry prompt}
\label{subsec:geometry}

In the G2F diffusion model sketched in Fig.~\ref{fig:DDPM}, we introduce the geometry prompt as the condition $\bs{c}$ in Eq.~\eqref{eq:reverse} and use a U-Net~\cite{Ronneberger2015} with a cross-attention mechanism~\cite{Rombach2022} to realize $\bs\epsilon_{\bs\theta}(z_t, t; \bs{c})$. 
The U-Net consists of a shrinking path, an expanding path, and some skip connections between the two. 
The shrinking path gradually compresses a low-dimensional, high-resolution field to a high-dimensional, low-resolution field. 
It contains five layers, with sizes of (128, 64, 64), (128, 32, 32), (256, 16, 16), (384, 8, 8), and (512, 4, 4), followed by an average pool layer to obtain encoded data with size of (512, 1, 1). 
The expanding path follows the reverse process. 

To incorporate the geometry prompt, we first encode the geometry information into a vector of 0 and 1, representing the fluid and solid, respectively. 
The prompt is encoded by an NN having the same size as the shrinking path. 
Outputs from each layer of the NN are then combined into the corresponding layers in the expanding path by attention mechanisms~\cite{Rombach2022}. 
We applied the geometric conditioning only in the expanding path of the U-Net for several reasons. 
The geometry information undergoes degradation during spatial compression in the shrinking path, making conditioning less effective at these stages~\cite{Huang2023,Liu2024IEEE}. 
Meanwhile, the expanding path leverages the skip connections to recover fine-grained details while properly incorporating the geometry constraints. 
Our test (not shown) demonstrates that conditioning only in the shrinking path produces poor results, with the model struggling to interpret the geometry prompts. 
Focusing conditional injection in the expanding path achieves a balance between computational efficiency and feature representation, avoiding redundant parameters.

Several conditioning strategies exist, each with distinct advantages and limitations. 
The channel concatenation method offers a simple implementation that merges condition and noise in the channel dimension at the U-Net entry point~\cite{Pan2025}. 
Although computationally efficient, it has limited modeling capacity for complex conditions and struggles with spatial alignment relations~\cite{Aladago2022}. 
Latent space condition injection encodes conditions before concatenation, reducing the computational cost for high-resolution generation~\cite{Rombach2022}. 
However, this requires a reliable auto-encoder and may lose details~\cite{Maheshwari2022}. 
Compared to other strategies, our implementation offers a balanced approach. 
It provides fine-grained control over feature interactions at specific resolutions, enabling precise geometric guidance without excessive computational burden.

Generative models produce diverse samples via stochastic latent space sampling. 
Conditional diffusion models use cross-attention to encode conditions, guiding the reverse diffusion process. 
Since DDPM starts from random noise, its results vary even with identical prompts. 
We focus here on geometry prompts, generating flow fields for a given geometry without specifying time. 
A model for generating time-specific flow fields is discussed in the Appendix~\ref{appendix:phase}.

\subsection{Datasets}\label{subsec:datasets}

\begin{figure}[ht]
  \centering
  \includegraphics[width=\textwidth]{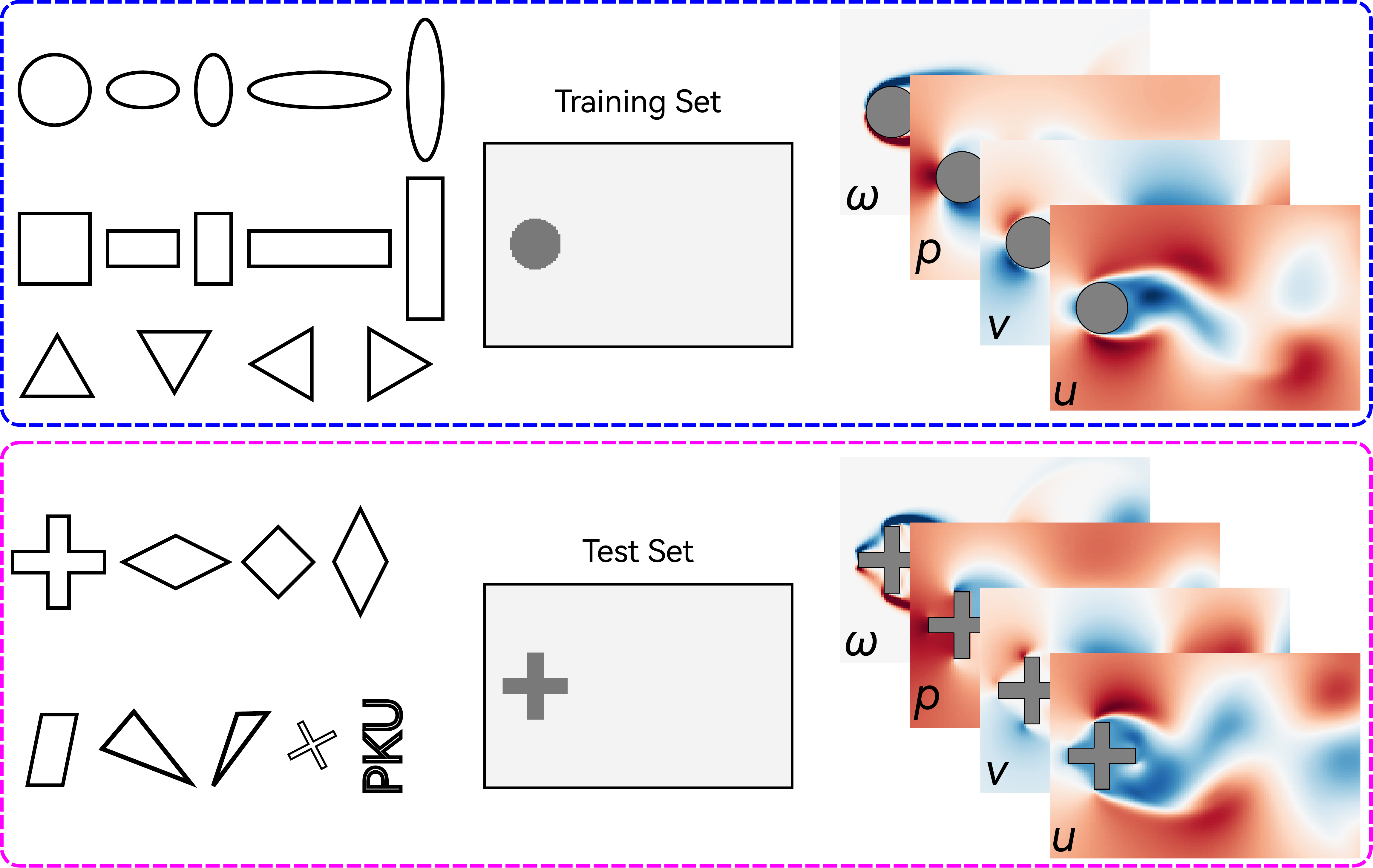}
  \caption{
  Datasets for evaluating the G2F diffusion model. 
  Elementary obstacle geometries including the circle, ellipse, rectangle, and triangle are used for training; 
  Complex obstacle geometries including the parallelogram, cross, and characters `\textsf{PKU}' are employed for testing. 
  Each sample $\bs{z}_0 = \lrs{u, v, p, \omega}$ includes velocity, pressure, and vorticity. 
  }
\label{fig:data}
\end{figure}

We evaluate the G2F diffusion model based on 2D flows around various obstacles. 
The dataset includes a range of obstacle geometries, as illustrated in Fig.~\ref{fig:data}. 
The G2F diffusion model is trained on the flow field around simple geometries, including the circle, ellipse, rectangle, and triangle. 
At each training step, we randomly select a geometry and one of the corresponding flow fields as input and target of the G2F diffusion model. 
To assess the model's ability to generalize to different geometries, the test set includes flows around obstacles not present in the training set. 
This includes several cases with triangle and parallelogram obstacles.
Furthermore, the model is tested on crosses and characters `\textsf{PKU}' with convex shapes.

The geometry complexity is characterized by the roundness $r = 4\pi S / C^2$, where $S$ and $C$ are the area and circumference, respectively. 
The roundness is bounded by $r=1$ for the circle and $r\rightarrow 0$ for extremely complex geometry. 
The crosses in the test set has $r = 0.165$ and $0.321$, smaller than all shapes in the training set, and the characters `\textsf{PKU}' have $r \approx 0$.
Although the roundness is not a perfect index for geometry complexity, it remains a practical option due to its simplicity and intuitive interpretation. 
Note that the roundness only serves as an indicator in results evaluation and does not affect the model training.

Focusing on the geometry condition, we evaluate the model using flow field data with the same free-stream velocity $U_\infty = 1$, viscosity $\mu = 1\times 10^{-2}$, and density $\rho=1$. 
The circle radius is 1, and the characteristic length of the obstacles ranges from 1 (a rectangle or ellipse with its longitudinal axis aligned in the streamwise direction) to 4 (a rectangle or ellipse with its longitudinal axis aligned in the spanwise direction). 
Correspondingly, the Reynolds number $\Rey$ varies from 100 to 400.
Our study restricts the Reynolds number to a small range in the laminar regime. 
This allows us to isolate and clearly evaluate the impact of geometric variations on flow patterns, avoiding the complexities of turbulent physics. 
This focused approach is essential for assessing our G2F model's response to different geometries. 
It provides a foundation for validating the model's generative capabilities concerning flow physics driven directly by streamwise geometry.

We solve the incompressible Navier-Stokes equations to obtain the flow field around various obstacles as the ground truth (GT) using OpenFOAM~\cite{Greenshields2022}.
The body-fitted meshes with 55000 cells are employed for a computational domain of $\lrr{-20, 60}\times\lrr{-20, 20}$, where the geometric centroid of the obstacle is positioned at the spatial location $(x,y) = (0,0)$. 
Each flow-field snapshot has the velocity $(u, v)$, pressure $p$, and vorticity $\omega$. 
Note that we interpolate the flow field to a $128\times 128$ uniform grid over a view box of $\lrr{-2, 10}\times\lrr{-4, 4}$ to obtain $z_0=[u, v, p, \omega]$ for the model. 
For each case, 50 to 60 snapshots are sampled from one period. 

We set $T = 800$ and $\beta_t = 1\times 10^{-4} + 2.5\times 10^{-5} t$ empirically. 
The parameters $\bs\theta$ in the NN are optimized to minimize the loss function in Eq.~\eqref{eq:lossMSE}.
One geometry and corresponding snapshots are taken from the training dataset randomly at each model training step. 
The parameters are initialized by the Kaiming method~\cite{Kaiming2015} and updated using the Adam optimizer~\cite{Adam2015}. 
The initial learning rate is 0.001, subsequently reduced to one-tenth of its preceding value after every 2000 epochs.

\subsection{Comparison with other models}

We evaluate our G2F diffusion model against deterministic and probabilistic baselines to assess its performance comprehensively. 
For direct architectural comparison, we implement two CNN models using the identical U-Net architecture: CNN-all (CNN-a), trained on the complete dataset across all geometries, and CNN-snapshot (CNN-s), trained on single snapshots per geometry that most closely match our diffusion model's outputs.

We also implement a VAE~\cite{Kingma2014} as a representative probabilistic generative model for comparison. 
This comparison helps evaluating how effectively our diffusion-based approach captures the distribution of possible flow fields.
The VAE architecture is adapted from the U-Net's shrinking path (encoder) and expanding path (decoder), ensuring comparable network complexity. 
The encoder $q_\phi \lrr{\mathbf{z}\vert\mathbf{x}}$ maps flow fields $\bs{x}$ into latent space parameters $\lrr{\mu, \sigma}$, which define a Gaussian distribution $\bs{z} = \mu + \sigma \odot \epsilon$, where $\epsilon \sim \mc{N}\lrr{0, 1}$.
The decoder $p_\theta \lrr{\mathbf{x}\vert\mathbf{z}}$ then reconstructs $\bs{z}$ to predicted flow yields $\tilde{\bs{x}}$. 
We employ the standard $\beta$-VAE~\cite{Higgins2017} loss function
\EQ
  \mc{L}_\text{VAE}(\phi, \beta) = - \mathbb{E}_{\mathbf{z} \sim q_\phi \lrr{\mathbf{z}\vert\mathbf{x}}} \log p_\theta \lrr{\mathbf{x}\vert\mathbf{z}} + \beta \cdot D_\text{KL}\lrr{ q_\phi \lrr{\mathbf{z}\vert\mathbf{x}}\|p_\theta\lrr{\mathbf{z}}},
\EN
where $D_\text{KL}$ is the KL divergence and $\beta = 0.1$. 
Although GANs~\cite{Goodfellow2014} represent another class of generative models, we selected VAE for its balanced approach to fidelity and diversity, and its more stable training characteristics than adversarial methods. 

All models were trained on Tesla V100 GPUs, with training times of 15.8 hours for the G2F diffusion model, 11.3 hours for CNN-a, 11.1 hours for CNN-s, and 3.4 hours for VAE. 
Inference measurements revealed a computational trade-off: the G2F diffusion model required 10 seconds per sample due to its iterative sampling process, while CNN models executed in 0.02 seconds, and VAE achieved the fastest inference at 0.01 seconds per sample.

\section{Results}\label{sec:results}

We evaluate the G2F diffusion model against the obstacles in the training and test sets.
Because the model relies on random noise input, we cannot guarantee a prediction at a specific time. 
Otherwise, we select a snapshot in the GT solution for comparison with the model output.
We measure the similarity between the model output and GT using the $L_1$ error 
\EQ\label{eq:L1}
    L_1\lrr{\bs{z}_0^M} =  \lrN{ \bs{z}_0^M - \bs{z}_0^{\mr{GT}} },
\EN
where the superscript $M$ denotes the model output. 
Then the GT snapshot with the minimal $L_1\lrr{\bs{z}_0^M}$ is used for comparison below in each geometry case, and these snapshots are selected to train the CNN-s.

\subsection{Training set}\label{subsec:train}

\begin{figure}[ht]
  \centering
  \begin{overpic}[height=0.3\textheight]{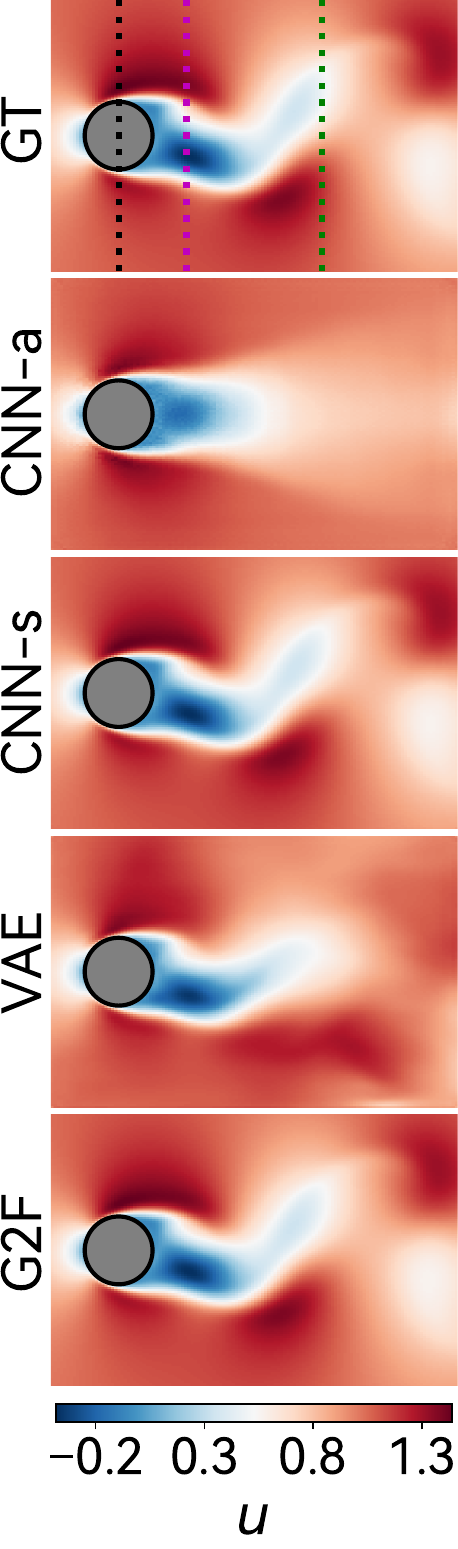}
    \put(-5, 96){(a)}
  \end{overpic}\hspace{15pt}
  \begin{overpic}[height=0.3\textheight]{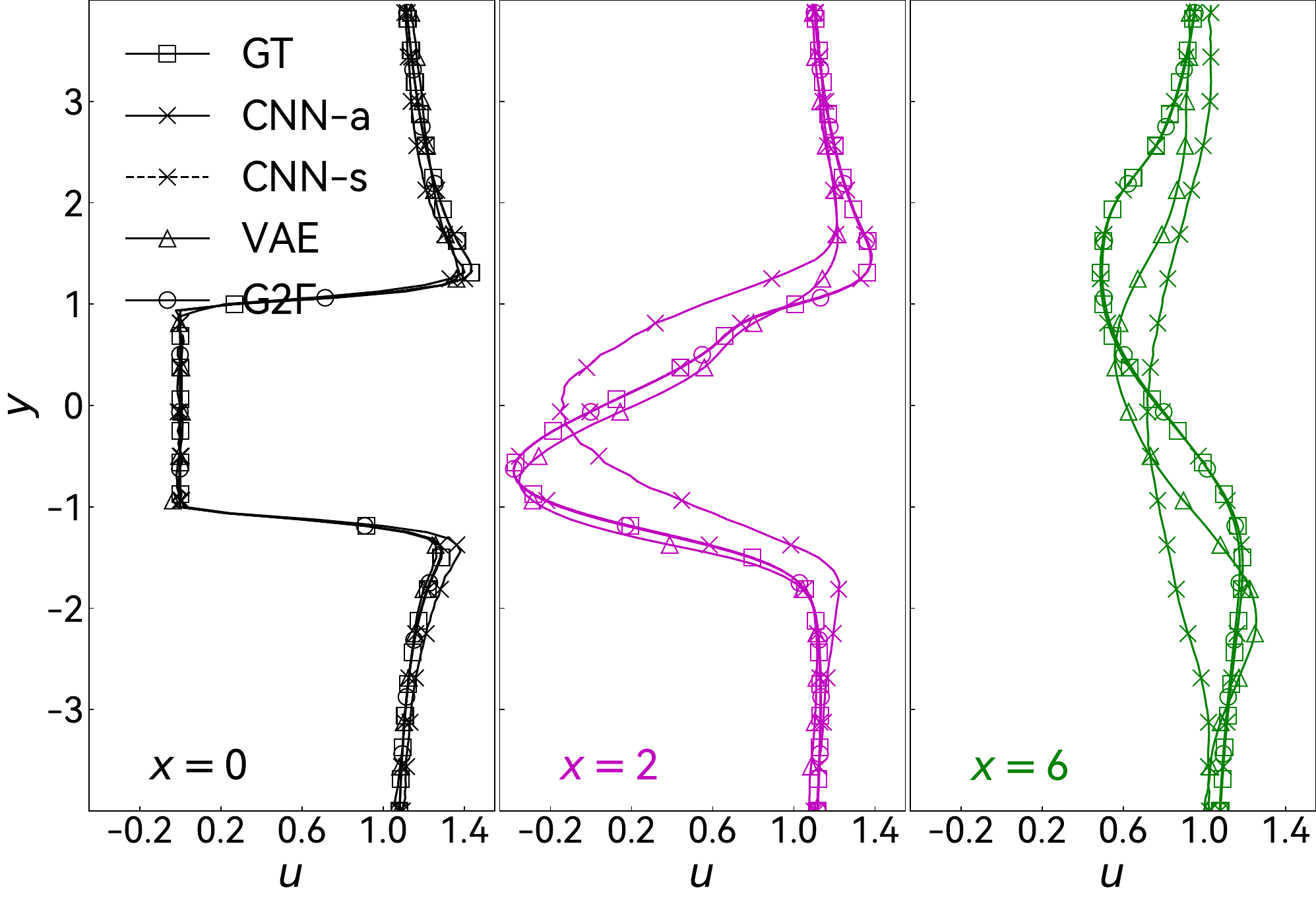}
    \put(0, 65){(b)}
  \end{overpic}
  \caption{
  Flow around a cylinder generated by the G2F, VAE, CNN-a, and CNN-s models, compared with the GT: (a) contour of $u$, (b) profiles of $u$ in the wake at $x=$ 0, 2, and 6.
  The vertical dashed lines in the upper left panel mark $x=$ 0, 2, and 6.
  }
\label{fig:circle}
\end{figure}

We first present the flow around a cylinder at $\Rey = 200$ as a benchmark ML-enhanced CFD case, characterized by periodic vortex shedding. 
Figure~\ref{fig:circle}(a) shows the instantaneous streamwise velocity $u$ contours, where G2F, VAE, and CNN-s successfully capture the von K\'{a}rm\'{a}n vortex street in the wake.
In contrast, CNN-a produces a ``time-averaged'' flow field with symmetric structures because its training minimizes the distance to all snapshots. 
Although CNN-s matches its specific training snapshot, it cannot reproduce other temporal states in the same period.
The VAE effectively models primary structures near the cylinder but loses accuracy in far-wake regions. 
Velocity profiles at $x = 0$, 2, and 6 in Fig.~\ref{fig:circle}(b) confirm these observations: 
G2F and CNN-s correctly predict the wake's evolving peaks and valleys; 
VAE performs well near the cylinder ($x = 0$ and 2) but degrades farther downstream ($x = 6$);
CNN-a yields unrealistic symmetric profiles.

\begin{figure}[ht]
  \centering
  \begin{overpic}[height=0.3\textheight]{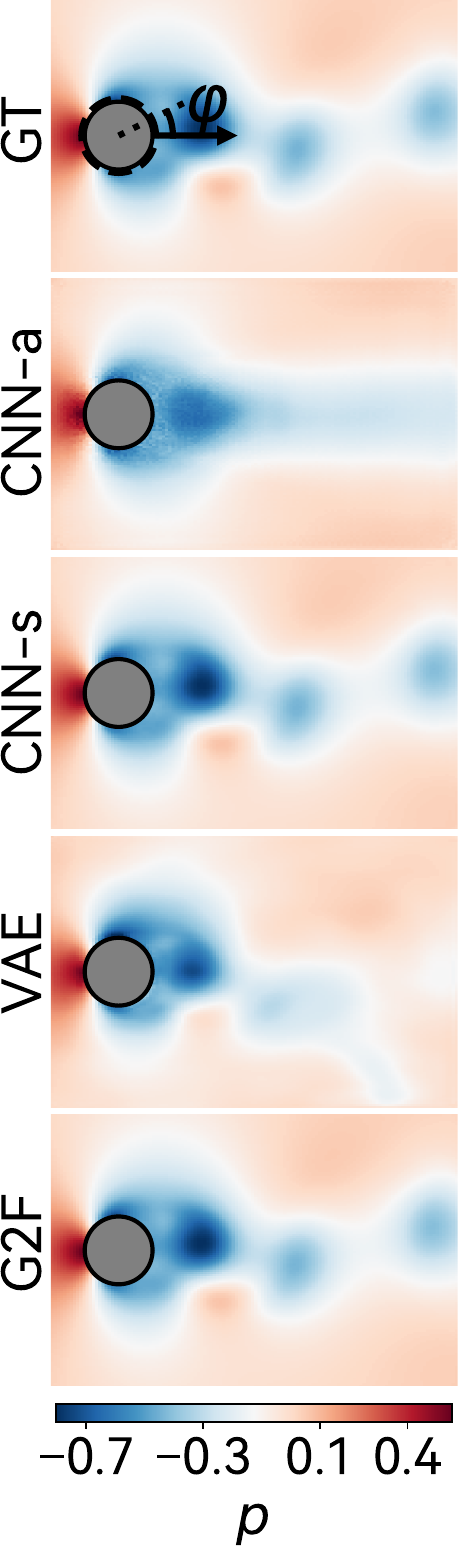}
    \put(-5, 96){(a)}
  \end{overpic}\hspace{15pt}
  \begin{overpic}[height=0.3\textheight]{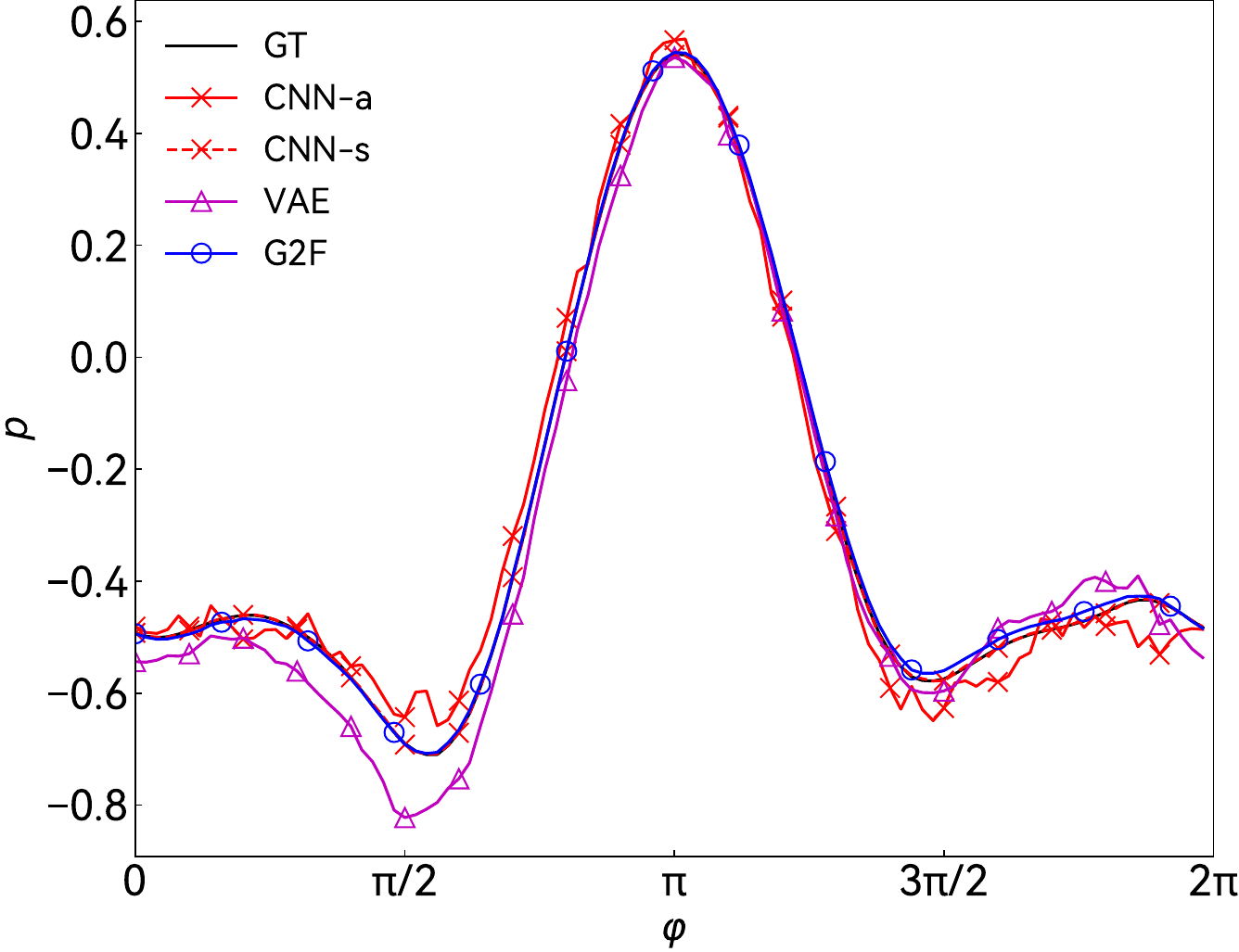}
    \put(0, 73.8){(b)}
  \end{overpic}
  \caption{
  Flow around a cylinder generated by the G2F, VAE, CNN-a, and CNN-s models, compared with the GT: (a) contour of pressure $p$, and (b) pressure on the obstacle surface. 
  }
\label{fig:circle_p}
\end{figure}

Figure~\ref{fig:circle_p} plots the instantaneous pressure distribution, showing (a) pressure contours and (b) surface pressure measurements at different angular positions $\varphi$.
All models successfully capture the high-pressure stagnation region at the cylinder's front, though with varying downstream accuracy. 
CNN-s, VAE, and G2F effectively approximate the alternating low-pressure zones, while CNN-a fails to represent this natural asymmetry. 
The surface pressure comparison in Fig.~\ref{fig:circle_p}(b) reveals more detailed differences: all models accurately predict the maximum pressure at the stagnation point ($\varphi = \pi$), but diverge in other regions. 
G2F demonstrates the most consistent agreement with GT across the entire cylinder surface, indicating a good prediction on the boundary layer separation. 
Meanwhile, VAE exhibits exaggerated pressure minima around $\varphi = \pi/2$ and $3\pi/2$. 
Both CNN models show reasonable agreement at the stagnation point but suffer from unphysical oscillations.

\begin{figure}[ht]
  \centering
  \begin{overpic}[height=0.3\textheight]{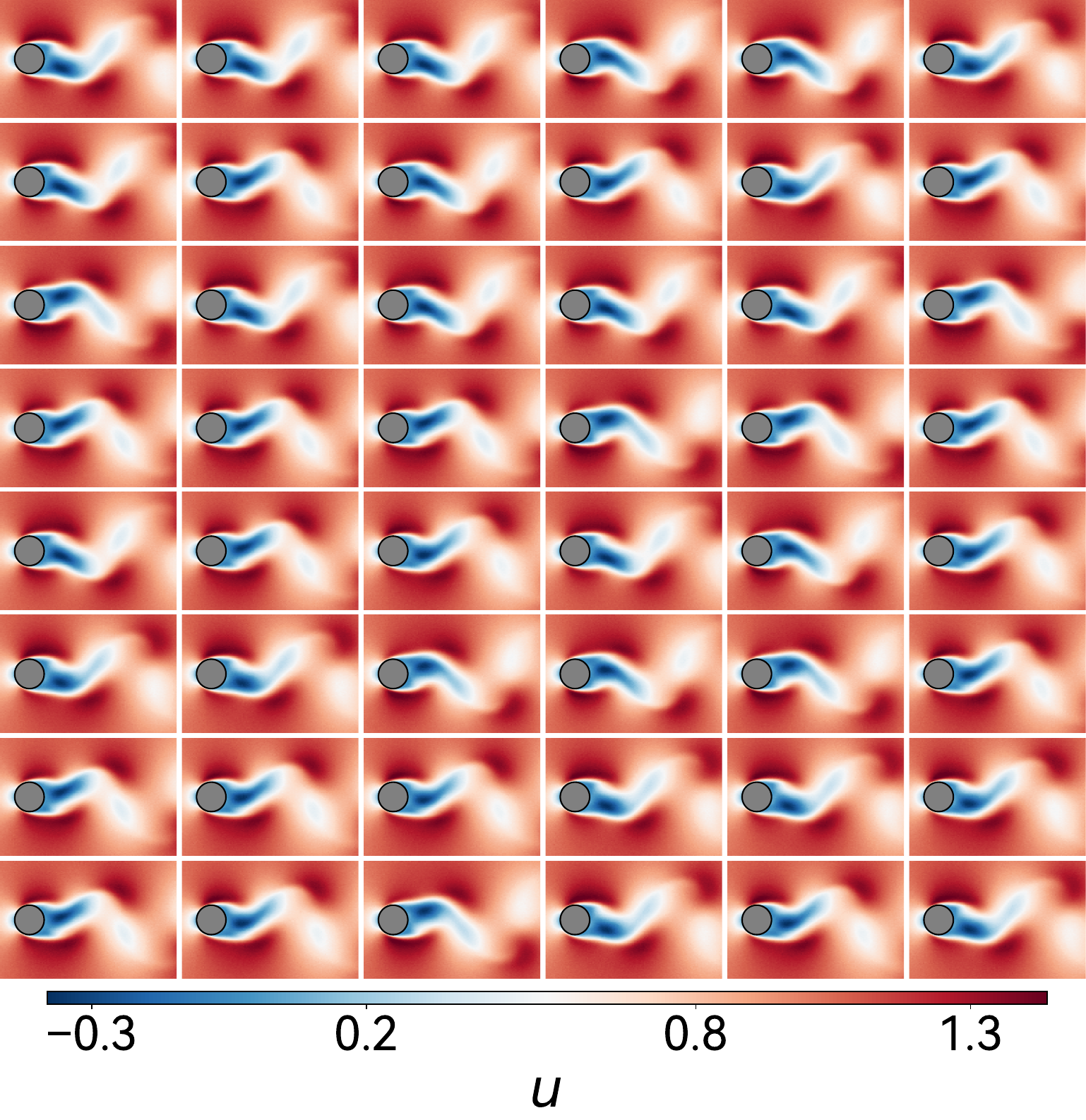}
    \put(-7, 95){(a)}
  \end{overpic}\hspace{15pt}
  \begin{overpic}[height=0.3\textheight]{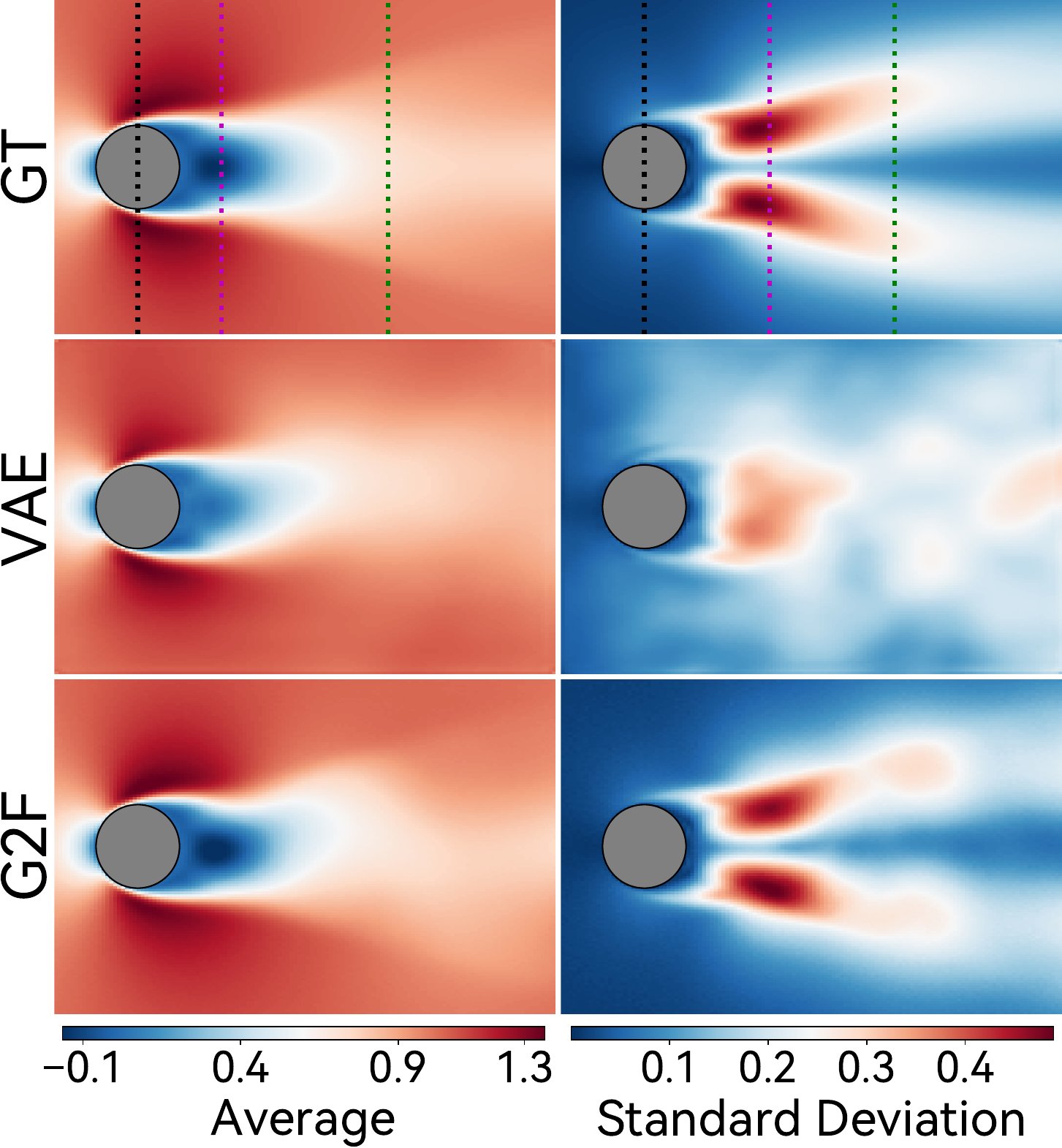}
    \put(-3, 95){(b)}
  \end{overpic}\\
  \begin{overpic}[width=0.47\textwidth]{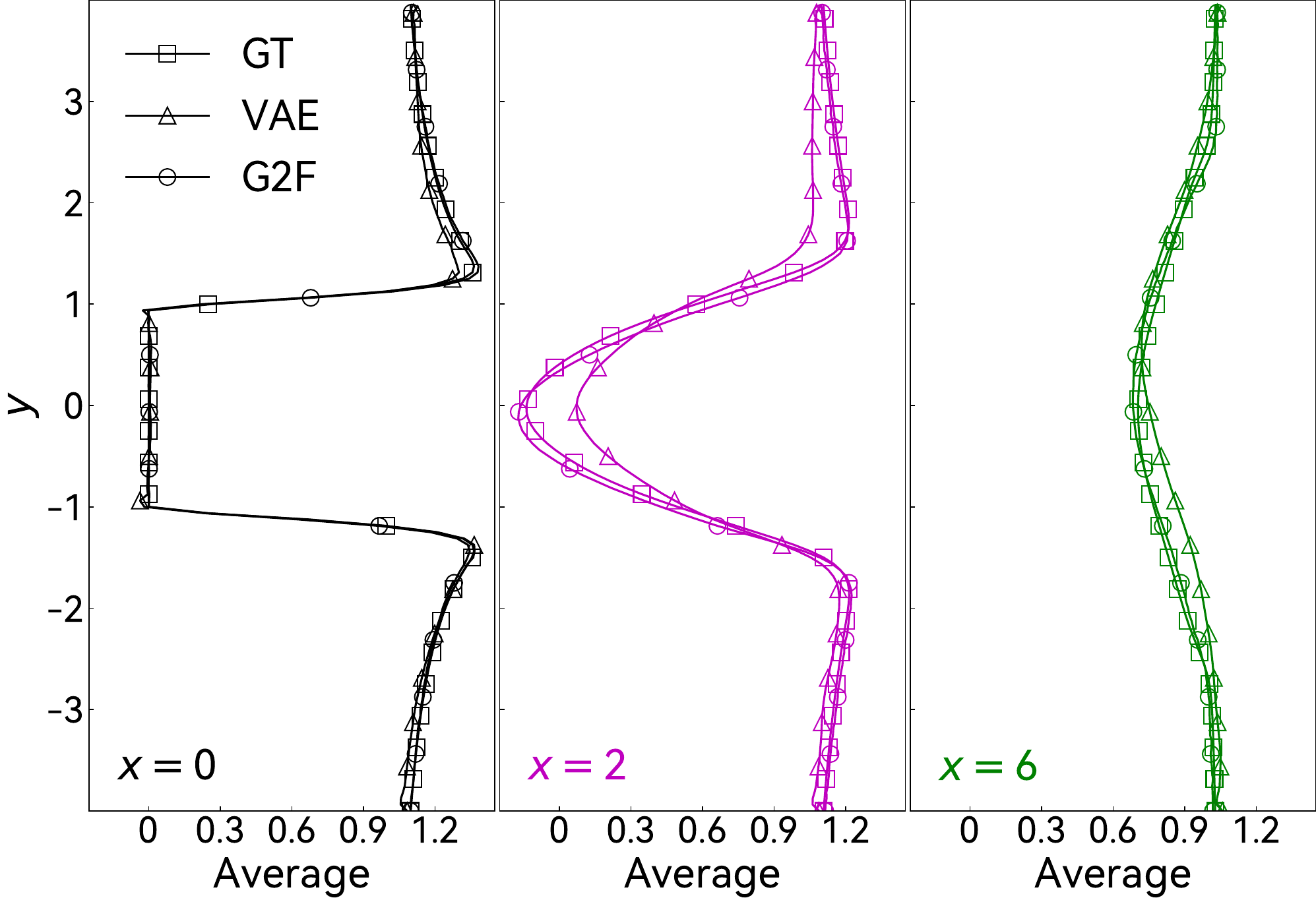}
    \put(-2, 63){(c)}
  \end{overpic}\hspace{15pt}
  \begin{overpic}[width=0.47\textwidth]{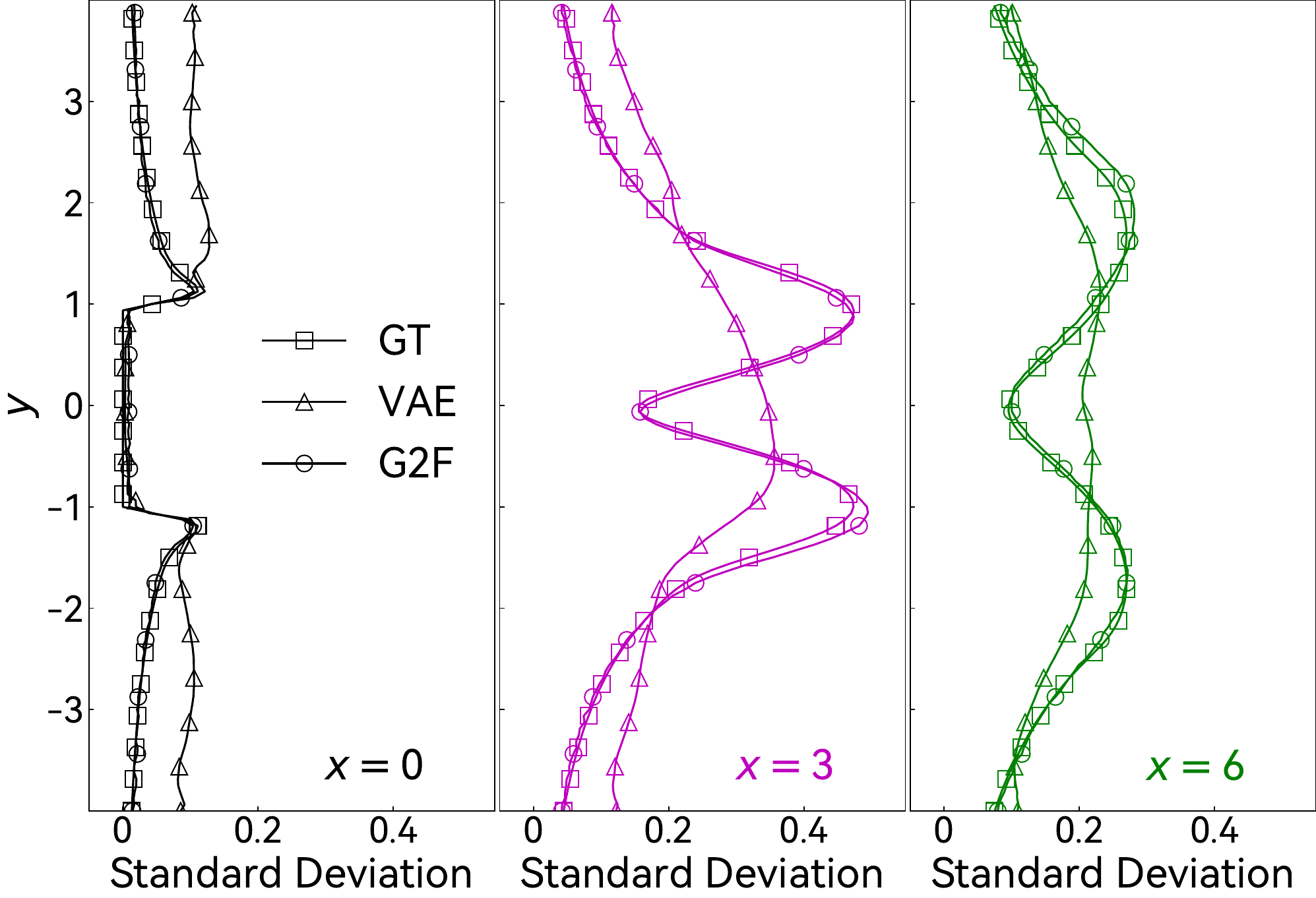}
    \put(-2, 63){(d)}
  \end{overpic}
  \caption{
    Statistical evaluation of generative models on flow around a cylinder:
    (a) generated samples of $u$ by the G2F diffusion model; 
    (b) comparison of mean and standard deviation of $u$ obtained via G2F, VAE, and GT. Statistics are computed over 50 generated samples;
    (c) mean streamwise velocity $u$ profiles at positions $x = $ 0, 2, and 6; 
    (d) standard deviation profiles of $u$ at $x = $ 0, 3, and 6.
  }
\label{fig:round-multi}
\end{figure}

Moreover, both diffusion and VAE models, being generative in nature, produce diverse samples from identical geometric prompts, which is a fundamental difference with deterministic CNN models. 
In flows past an obstacle, multiple flow fields can satisfy the same boundary conditions, particularly in regimes with vortex shedding and wake instabilities. 
The variance observed across the generated samples represents the model's exploration of the solution space for the given geometric conditions. 
This variance should be mainly interpreted as the model capturing the inherent distribution of possible flow states rather than uncertainty in the present work. 
The generated results shown in Fig.~\ref{fig:round-multi}(a) confirms that the G2F diffusion model captures this distribution of the time-varying flow phenomena.

By generating diverse samples that satisfy the same geometric constraints, the generative models enable exploration of multi-modal solutions in complex flow regimes and statistical analysis of flow fluctuation through ensemble averages.
Figure~\ref{fig:round-multi}(b) presents a statistical comparison of $u$ generated by G2F, VAE, and GT on the flow around a cylinder, where both the average flow field (left column) and standard deviation (right column) are compared with GT.
The G2F model outperforms the VAE in reproducing both mean flow structures and their fluctuation patterns. 
While both generative approaches capture the general average field characteristics, the G2F model reconstructs the twin high-variance regions in the cylinder wake, a signature feature indicating periodic vortex shedding. 
This reproduction of standard deviation patterns demonstrates that the G2F model better captures the time-varying flow phenomena.
Additionally, we include plots of the average and standard deviation of flow variables at different streamwise locations in Figs.~\ref{fig:round-multi}(c) and (d), respectively. 
Both G2F and VAE represent mean streamwise velocity profiles in the wake region, though the VAE shows a velocity deficit at $x = 2$. 
The G2F diffusion model precisely matches the GT standard deviation profile across all positions, capturing both vortex shedding peaks, while the VAE fails to reproduce this dual-peak structure.

Overall, the models exhibit distinct performance. 
The G2F diffusion model excels in reproducing instantaneous flow fields, capturing vortex structures, pressure distributions, and the statistics of vortex shedding, thanks to its generative nature and likelihood-maximizing training. 
The VAE predicts near-cylinder structures well but loses accuracy further downstream, showing exaggerated pressure minima and failing to capture the vortex shedding, a consequence of its limited latent space and lack of iterative refinement. 
CNN-s accurately reproduces its specific training snapshots but fails to generalize temporally. 
In contrast, CNN-a produces time-averaged, symmetric flow fields due to its objective of minimizing error across all snapshots. 
These differing behaviors stem from their distinct mathematical formulations and training objectives.

\subsection{Test set}\label{subsec:test}

Having tested the models' performance on training data, we now assess their generalization to unseen test configurations. 
We employ obstacle geometries absent from the training, including parallelograms, triangles, crosses, and the characters `\textsf{PKU}'.
While the parallelograms and triangles exhibit similarities to the simple geometries in the training set, the crosses and characters `\textsf{PKU}' serve as out-of-distribution cases, featuring combinations of elementary geometries and convex shapes. 

\begin{figure}[ht]
  \centering
  \begin{overpic}[height=0.3\textheight]{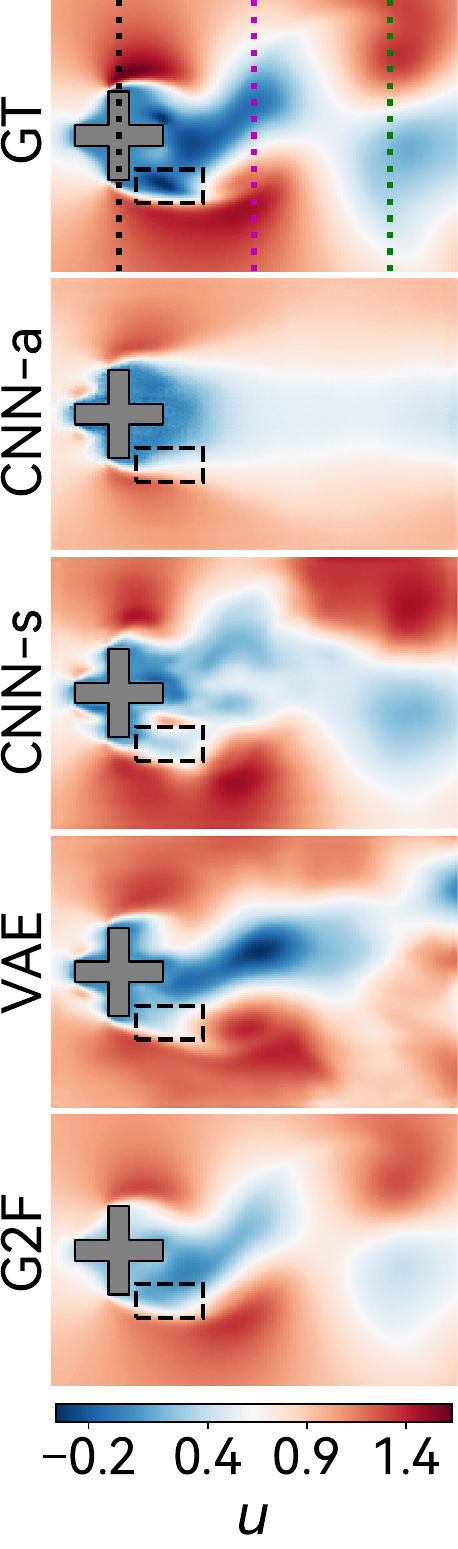}
    \put(-5, 96){(a)}
  \end{overpic}\hspace{15pt}
  \begin{overpic}[height=0.3\textheight]{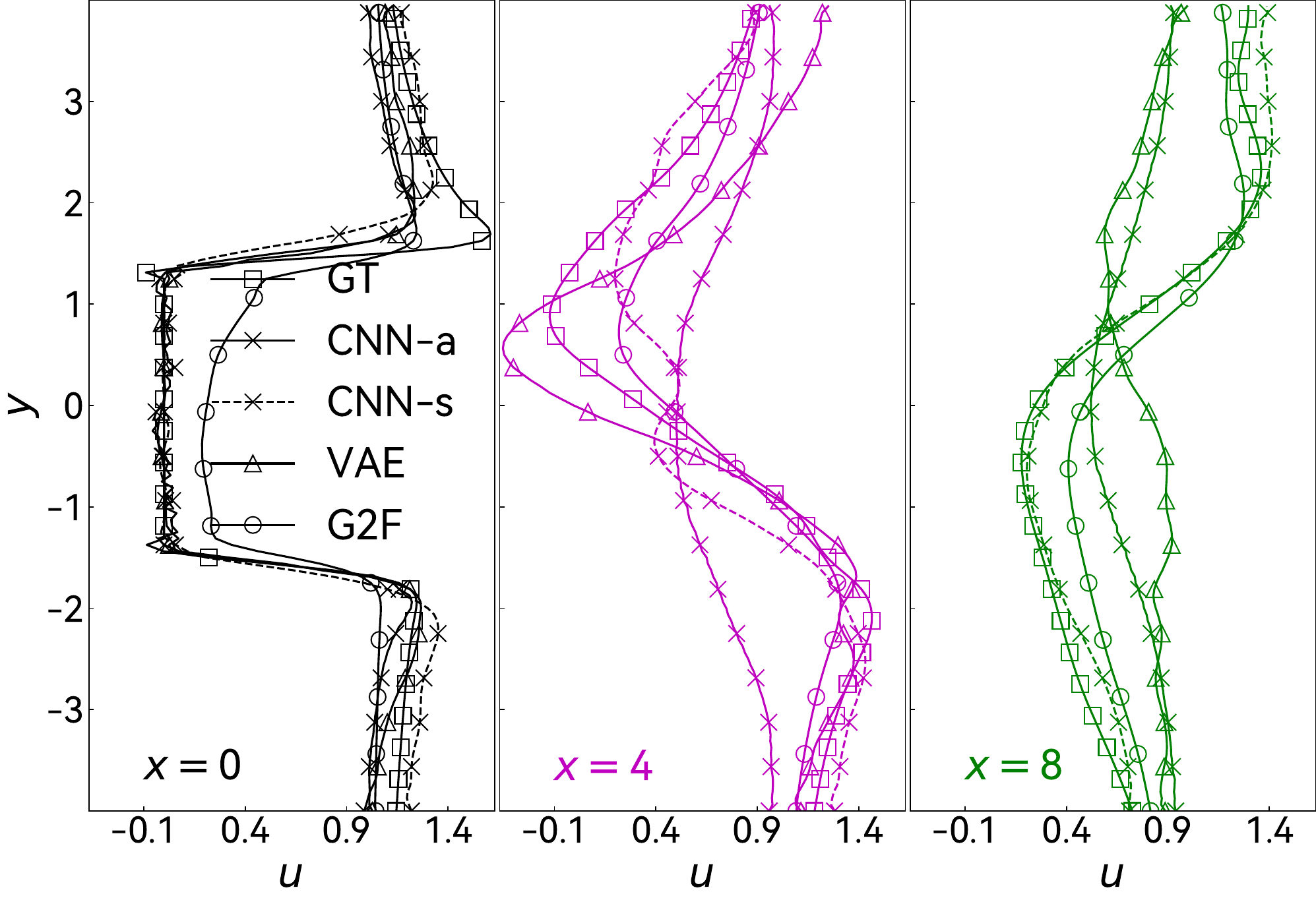}
    \put(0, 65){(b)}
  \end{overpic}
  \caption{
  Flow around a cross ($r = 0.321$) generated by the G2F, VAE, CNN-a, and CNN-s models, compared with GT: (a) contour of $u$, (b) profiles of $u$ in the wake at $x=$ 0, 4, and 8.
  The vertical dashed lines in the upper left panel mark $x=$ 0, 4, and 8.
  }
\label{fig:cross}
\end{figure}

Figure~\ref{fig:cross} compares the velocity field predictions across models for a cross-shaped obstacle with $r = 0.321$. 
The G2F diffusion model generates realistic flow fields characterized by shedding vortices and boundary layers, though it produces a narrower wake than GT. 
This limitation can be attributed to two factors: the attention mechanism's prioritization of near-obstacle regions~\cite{Vaswani2017} and the absence of geometries with large y-direction widths in the training set.
The velocity profiles at different x-positions highlight complex flow structures, including local $u$ minimums near $(x, y) = (1, -1.5)$ (marked in Fig.~\ref{fig:cross}(a)) caused by the cross's rear cavities.
This is a phenomenon absent from training geometries. 
The G2F diffusion model indicates a weak shear layer at this position, although it cannot accurately reproduce this structure. 
In contrast, the CNN-based models struggle with the velocity field predictions. 
The CNN-a model exhibits oscillations with averaged wake, while CNN-s produces fragmented and physically unrealistic patterns. 
The VAE model shows only marginally improvement over CNN-s in predicting the velocity field.

\begin{figure}[ht]
  \centering
  \begin{overpic}[height=0.3\textheight]{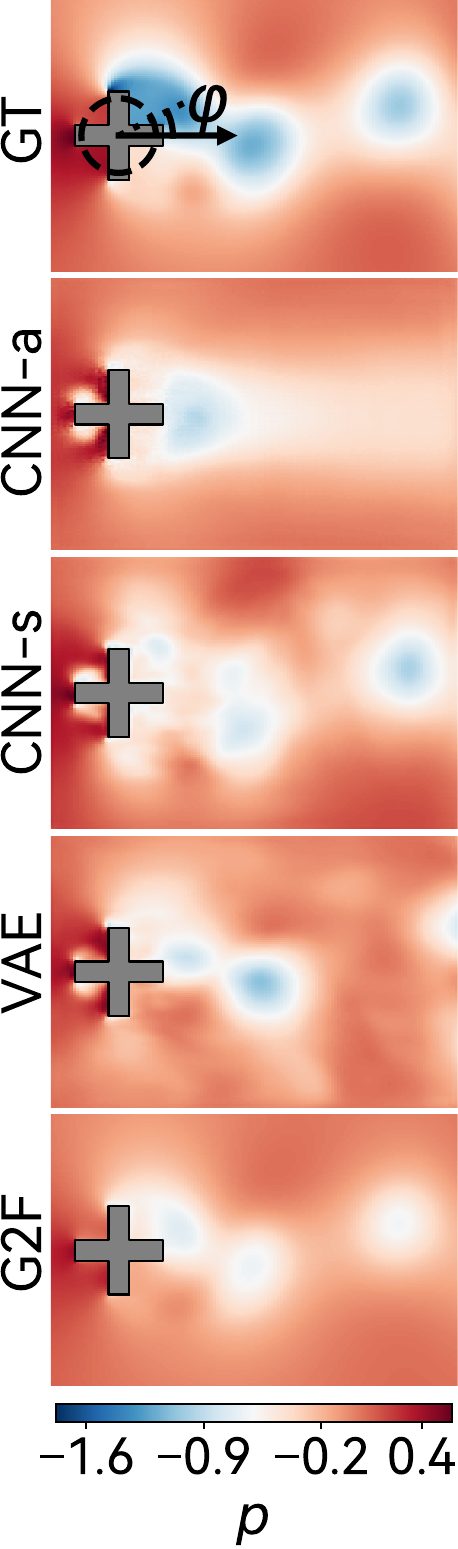}
    \put(-5, 96){(a)}
  \end{overpic}\hspace{15pt}
  \begin{overpic}[height=0.3\textheight]{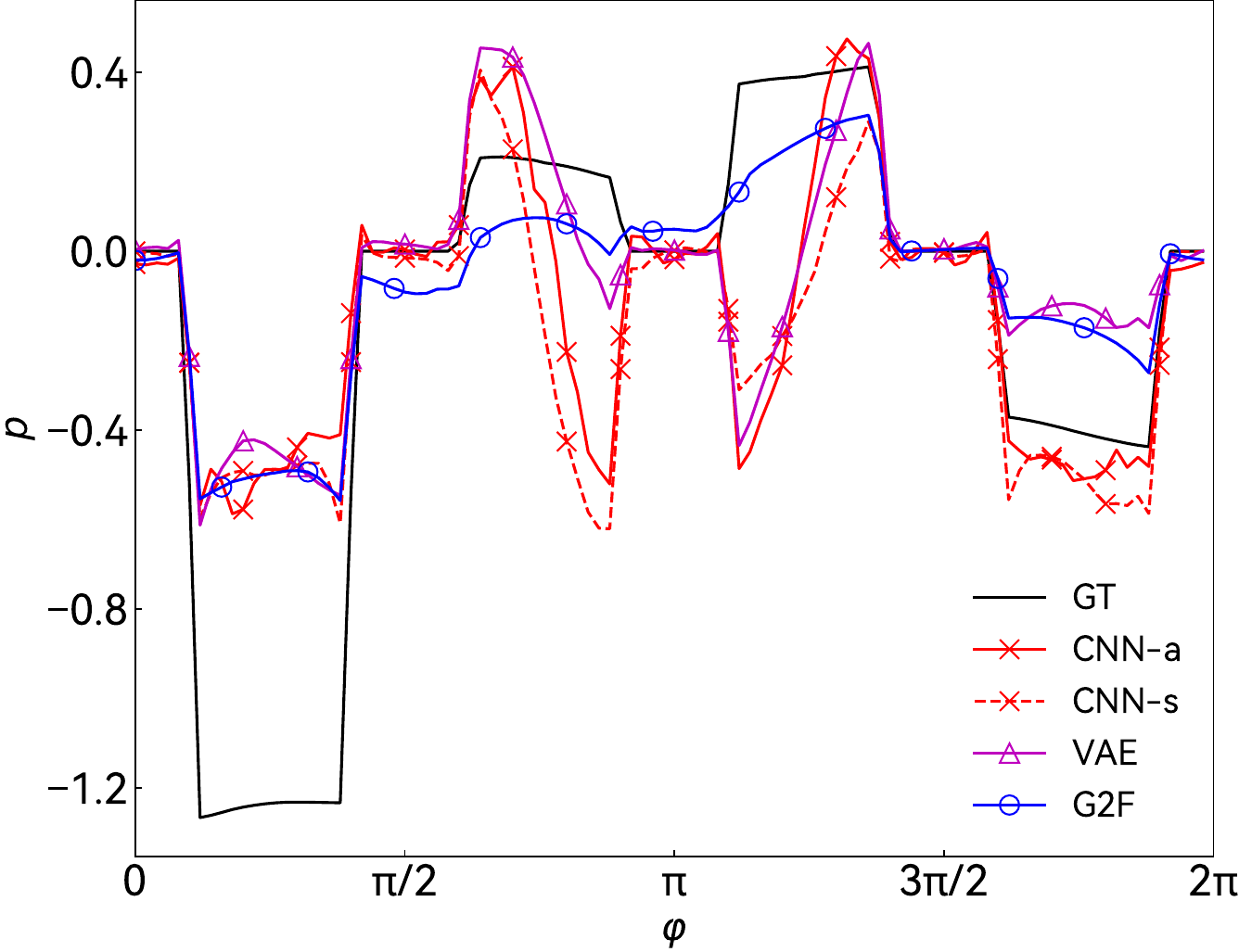}
    \put(0, 73.8){(b)}
  \end{overpic}
  \caption{
  Flow around a cross ($r = 0.321$) generated by the G2F, VAE, CNN-a, and CNN-s models, compared with GT: (a) contour of pressure $p$, and (b) pressure on a unit circle crossing the four arms. 
  }
\label{fig:cross_p}
\end{figure}

The pressure field comparisons in Fig.~\ref{fig:cross_p} further differentiate model performance.
The G2F model accurately captures the stagnation zones on the upstream side of the cross, as evidenced by the high-pressure values on the obstacle surface. 
The angular pressure distribution shown in Fig.~\ref{fig:cross_p}(b) reveals that G2F predictions follow the GT pattern most closely, particularly around key transition points.
Other models incorrectly predict two negative pressure regions near the front arm of the cross, suggesting that they treat the cross arms as standalone objects within the flow field.
This confirms that CNN-based and VAE models are insufficient for handling unseen geometries.
Despite its limitations, the G2F diffusion model effectively addresses the challenge of generalizing to unseen geometries, producing more physically coherent velocity and pressure fields than the other models.

\begin{figure}[ht]
  \centering
  \begin{overpic}[height=0.3\textheight]{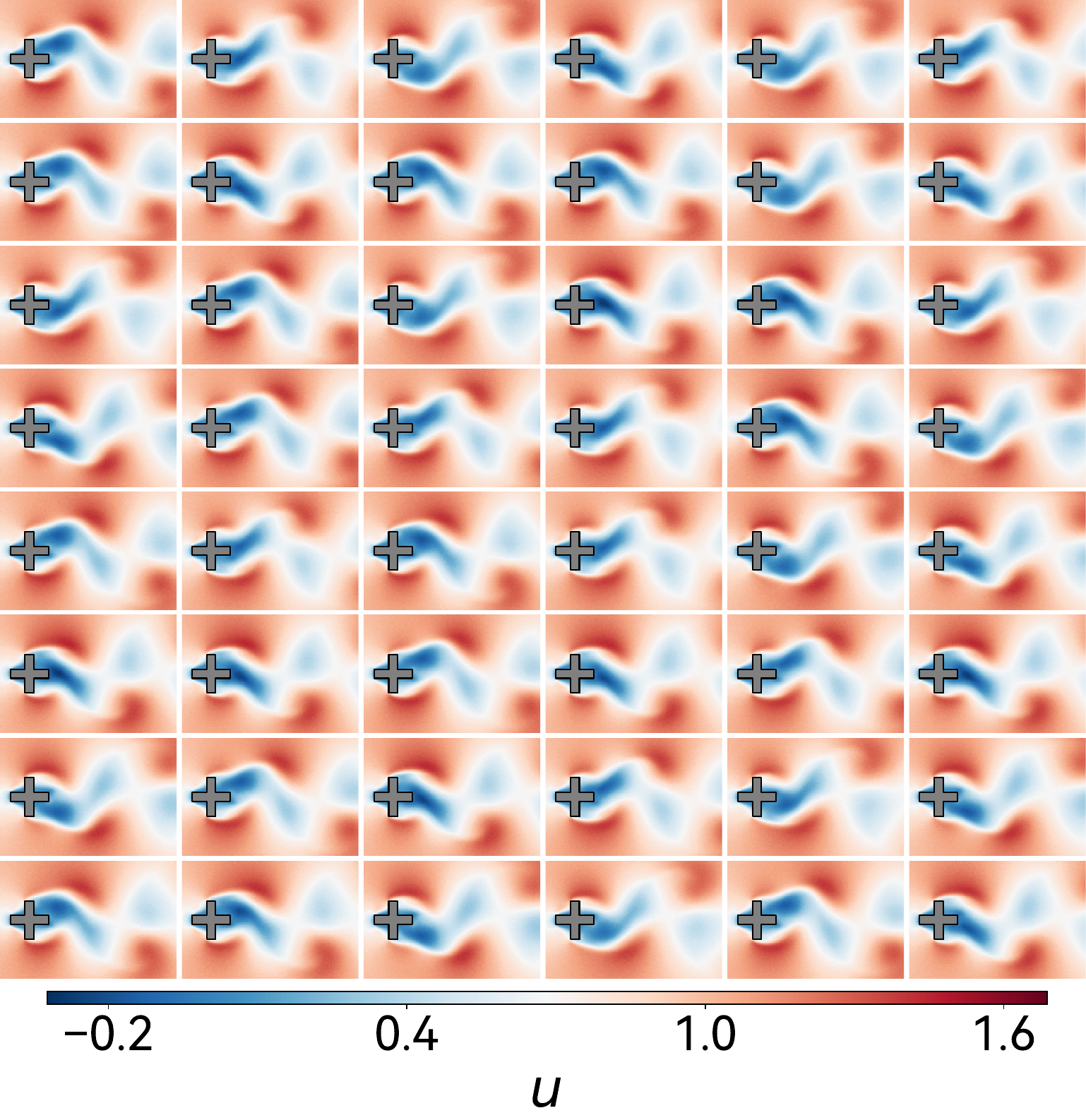}
    \put(-7, 95){(a)}
  \end{overpic}\hspace{15pt}
  \begin{overpic}[height=0.3\textheight]{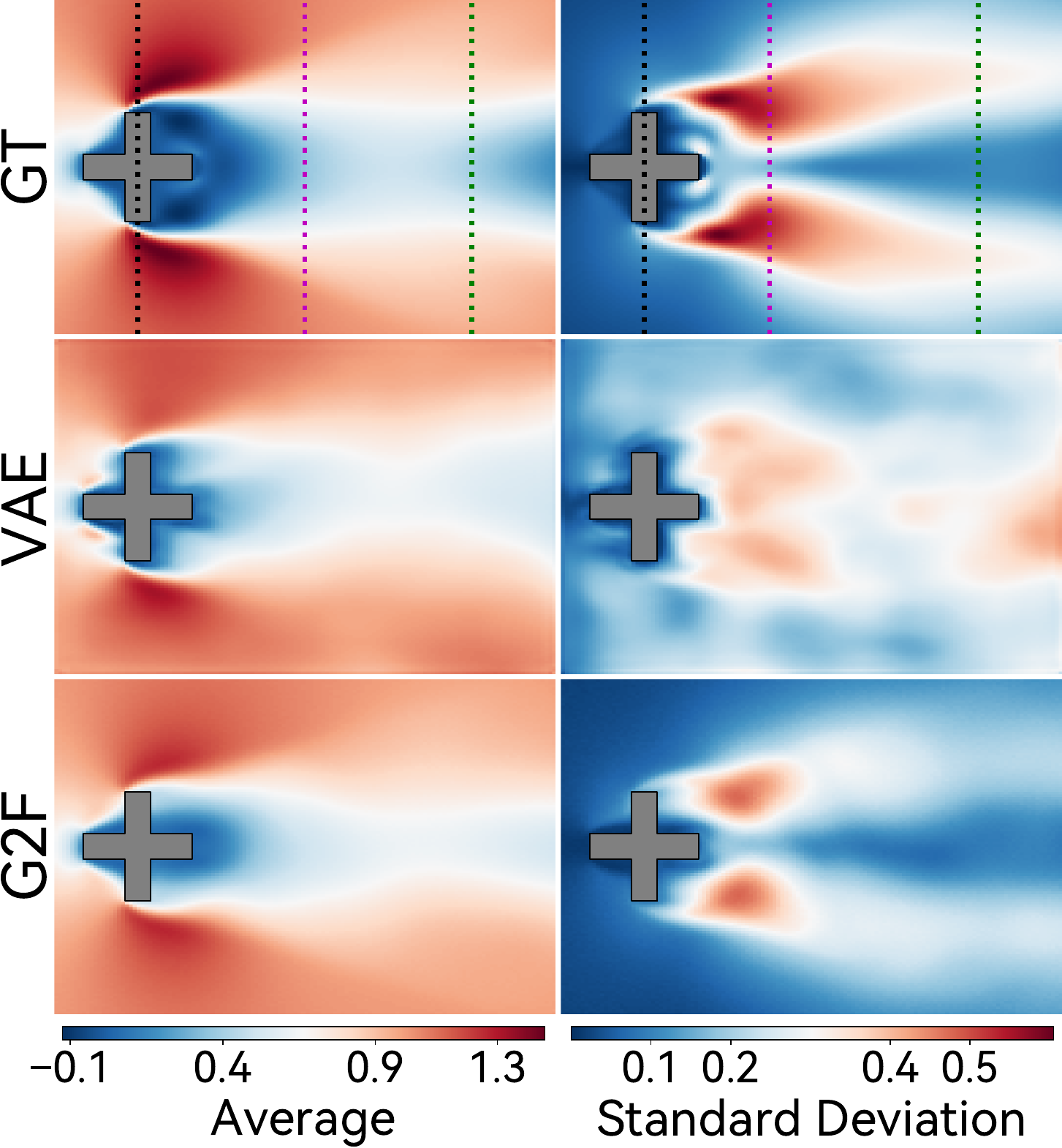}
    \put(-3, 95){(b)}
  \end{overpic}\\
  \begin{overpic}[width=0.47\textwidth]{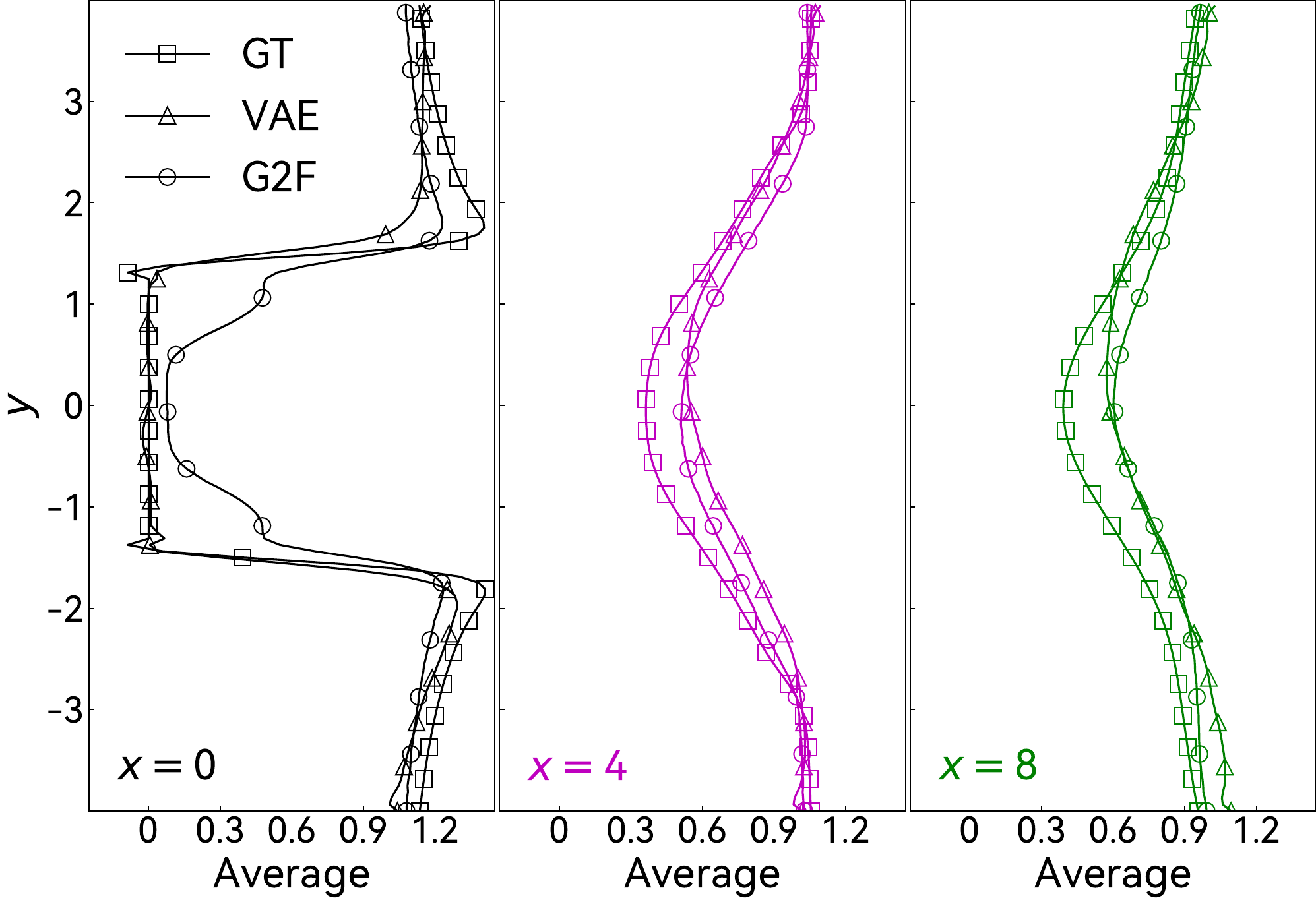}
    \put(-2, 63){(c)}
  \end{overpic}\hspace{15pt}
  \begin{overpic}[width=0.47\textwidth]{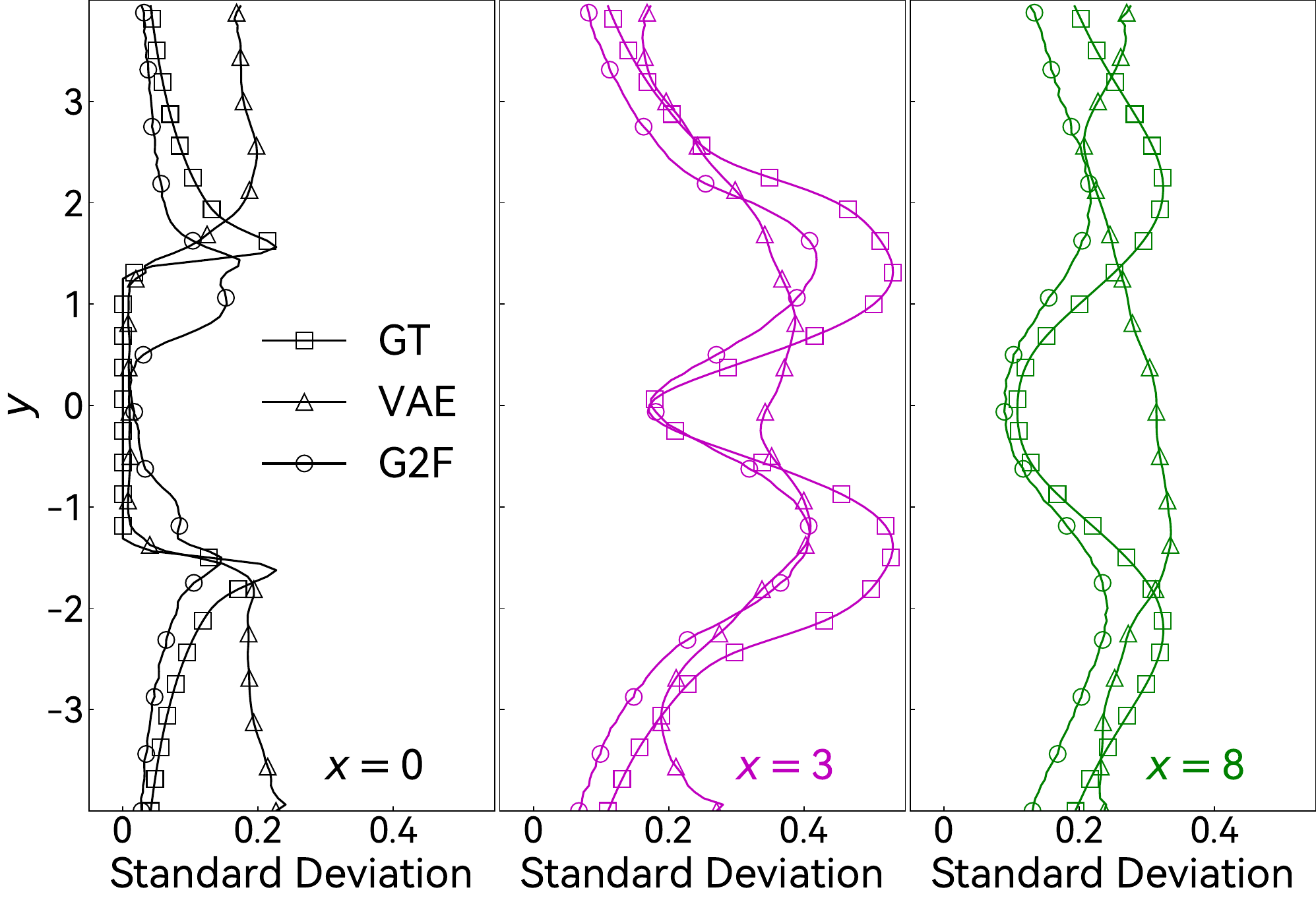}
    \put(-2, 63){(d)}
  \end{overpic}
  \caption{
    Statistical evaluation of generative models on flow around a cross ($r = 0.321$):
    (a) generated samples of $u$ by the G2F diffusion model; 
    (b) comparison of mean and standard deviation of $u$ among G2F, VAE, and GT. Statistics are computed over 50 generated samples;
    (c) mean streamwise velocity $u$ at $x = $ 0, 4, and 8;
    (d) standard deviation of $u$ at $x = $ 0, 3, and 8.
  }
\label{fig:cross-multi}
\end{figure}

Figure~\ref{fig:cross-multi} presents statistics of flow predictions for the cross case. 
The G2F diffusion model outperforms the VAE model across metrics, particularly in standard deviation field accuracy. 
However, both models exhibit certain limitations compared to GT. 
The G2F model generates average velocity fields that are narrower than GT, as previously noted in comparisons with CNN-based models. 
Additionally, G2F produce a smaller high standard deviation region than that in GT, suggesting incomplete capture of temporal variations. 
Quantitative comparisons of streamwise velocity profiles in Figs.~\ref{fig:cross-multi}(c) and (d) further substantiate these observations. 
Near the obstacle boundary at $x = 0$, G2F shows discrepancies that can be rectified through geometry matrix post-processing.
Throughout other regions, G2F reproduces velocity field structural features more accurately than VAE, successfully generating the characteristic periodic wake structure that matches GT's pattern. 
This demonstrates G2F's superior capability in reproducing complex flow statistics even with geometries outside the training distribution.

\begin{figure}[ht]
  \centering
  \includegraphics[width=\textwidth]{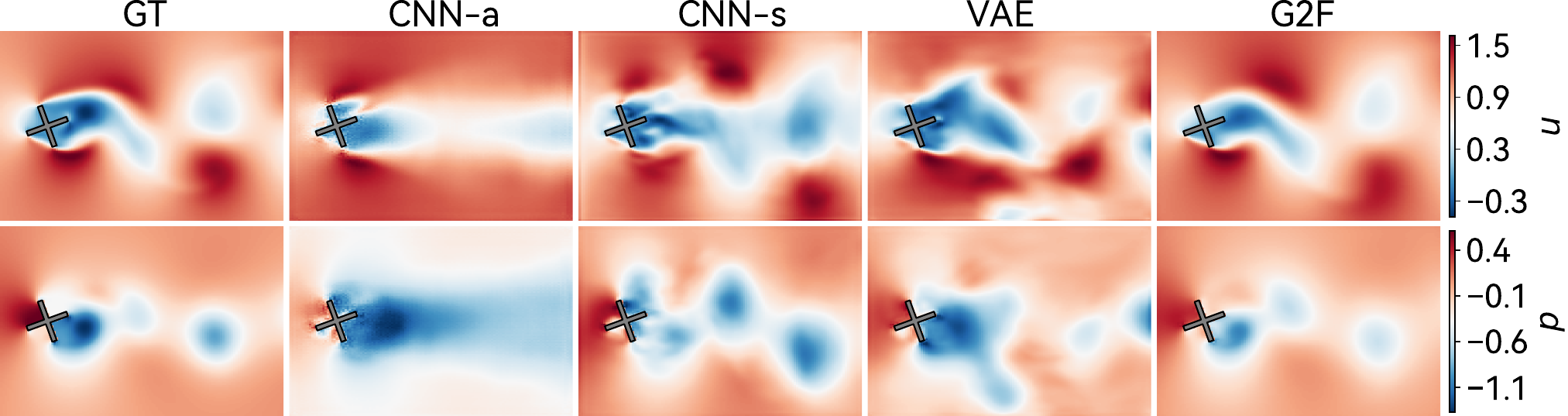}
  \caption{
    Flow around a rotated cross ($r = 0.165$) generated by the G2F, CNN-a, CNN-s, and VAE models, compared with GT, on contours of $u$ and $p$.
  }
\label{fig:cross2}
\end{figure}

Although generating a case with $r = 0$ is impossible, we include the generated results of the rotated cross for validation, which is much closer to zero roughness than the previous test. 
The rotated cross ($r = 0.165$) with its sharp features clearly distinguishes model performance in Fig.~\ref{fig:cross2}. 
G2F most accurately captures the flow separation, pressure gradients, and overall velocity and pressure distributions. 
In contrast, other models falter: CNN-a overly smooths features; CNN-s produces unphysical artifacts, incorrectly treating cross arms as independent; and VAE, though marginally better than CNN-s, shows errors, particularly in pressure field gradients and wake structures.

\begin{figure}[ht]
  \centering
  \includegraphics[width=0.85\textwidth]{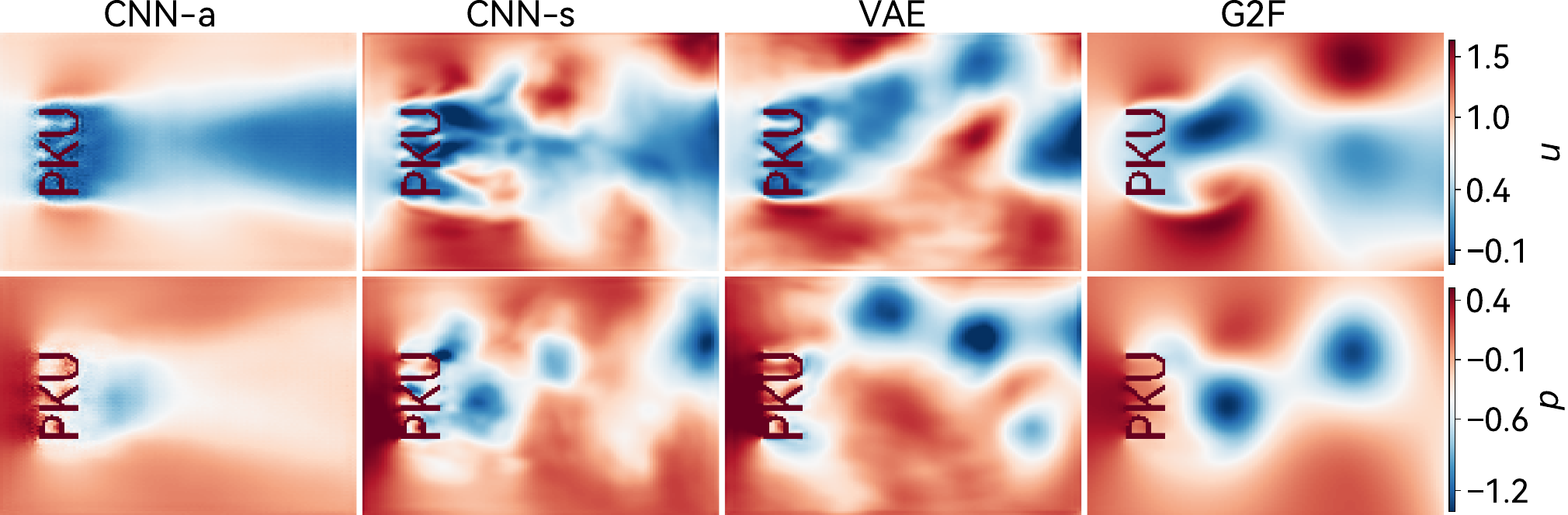}
  \caption{
    Flow around characters `\textsf{PKU}' generated by the G2F, CNN-a, CNN-s, and VAE models, on contours of $u$ and $p$.
  }
\label{fig:characters}
\end{figure}

Generating flow fields around the characters `\textsf{PKU}' is challenging due to the sharp corners, convex curves, and separated components, none of which are included in the training set. 
Given the difficulty in simulating this flow with a simple mesh, we present the generated streamwise velocity $u$ and pressure $p$ by the three models in Fig.~\ref{fig:characters}, without GT. 
The G2F diffusion model demonstrates remarkable generalization capability, effectively capturing the major flow features, despite some artifacts that make the flow field resemble the results for a rectangle. 
In contrast, the CNN-a model introduces noise at character boundaries and produces time-averaged-like fields. The CNN-s and VAE models generates a distinctly unphysical flow field.

Our comparison demonstrates that while both generative approaches can predict flow statistics, the G2F diffusion model's denoising architecture enables it to capture more complex flow structures and their temporal variations. 
Unlike the VAE, which directly maps to and from a single latent space, the diffusion model's step-by-step denoising mechanism effectively operates as a deeply hierarchical generative process~\cite{Yang2023}. 
This advantage becomes particularly evident in the standard deviation fields, where the diffusion model preserves more of the physically meaningful structures that characterize the vortex shedding.

\begin{figure}[ht]
  \centering
  \begin{overpic}[height=0.3\textheight]{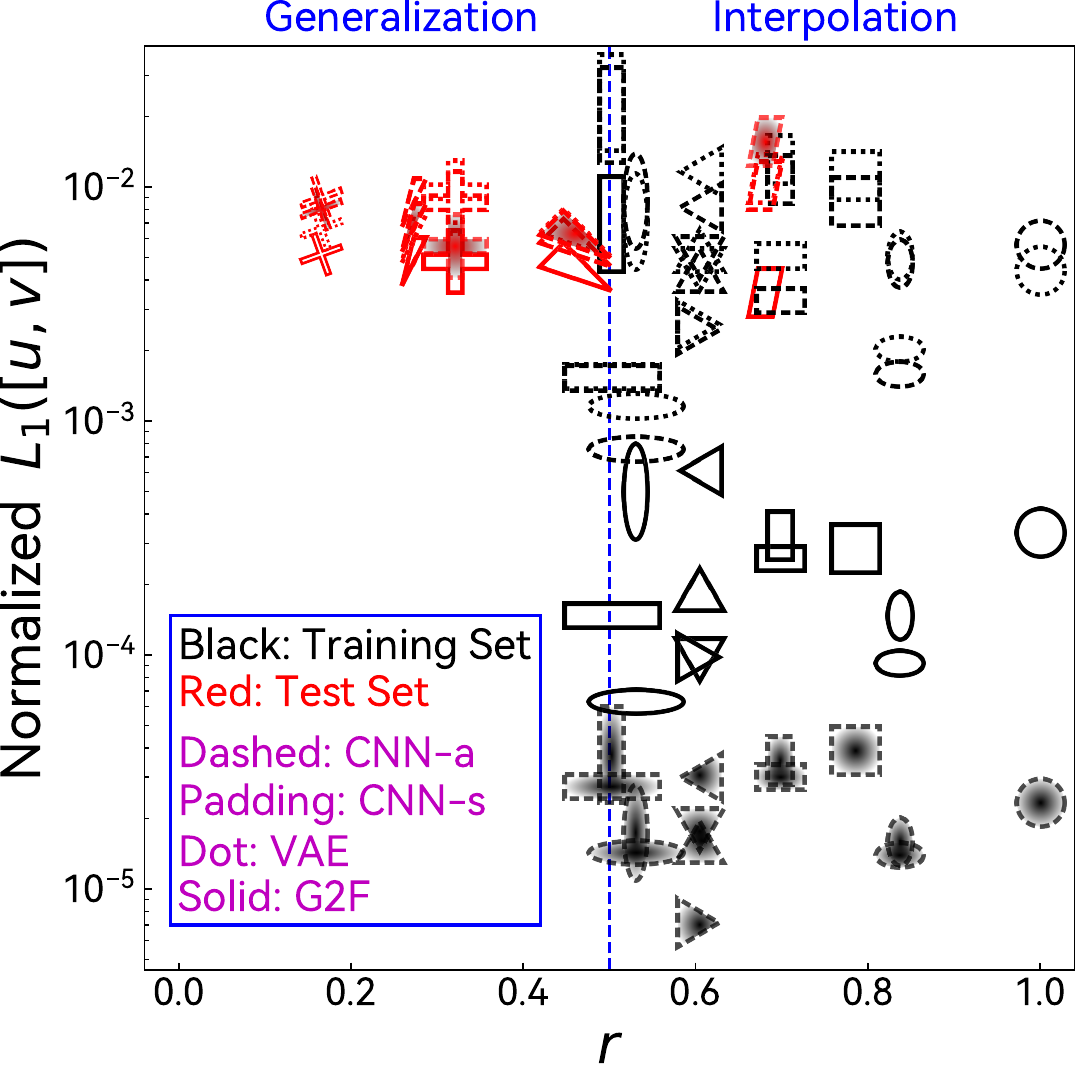}
    \put(6, 95){(a)}
  \end{overpic}\hspace{15pt}
  \begin{overpic}[height=0.3\textheight]{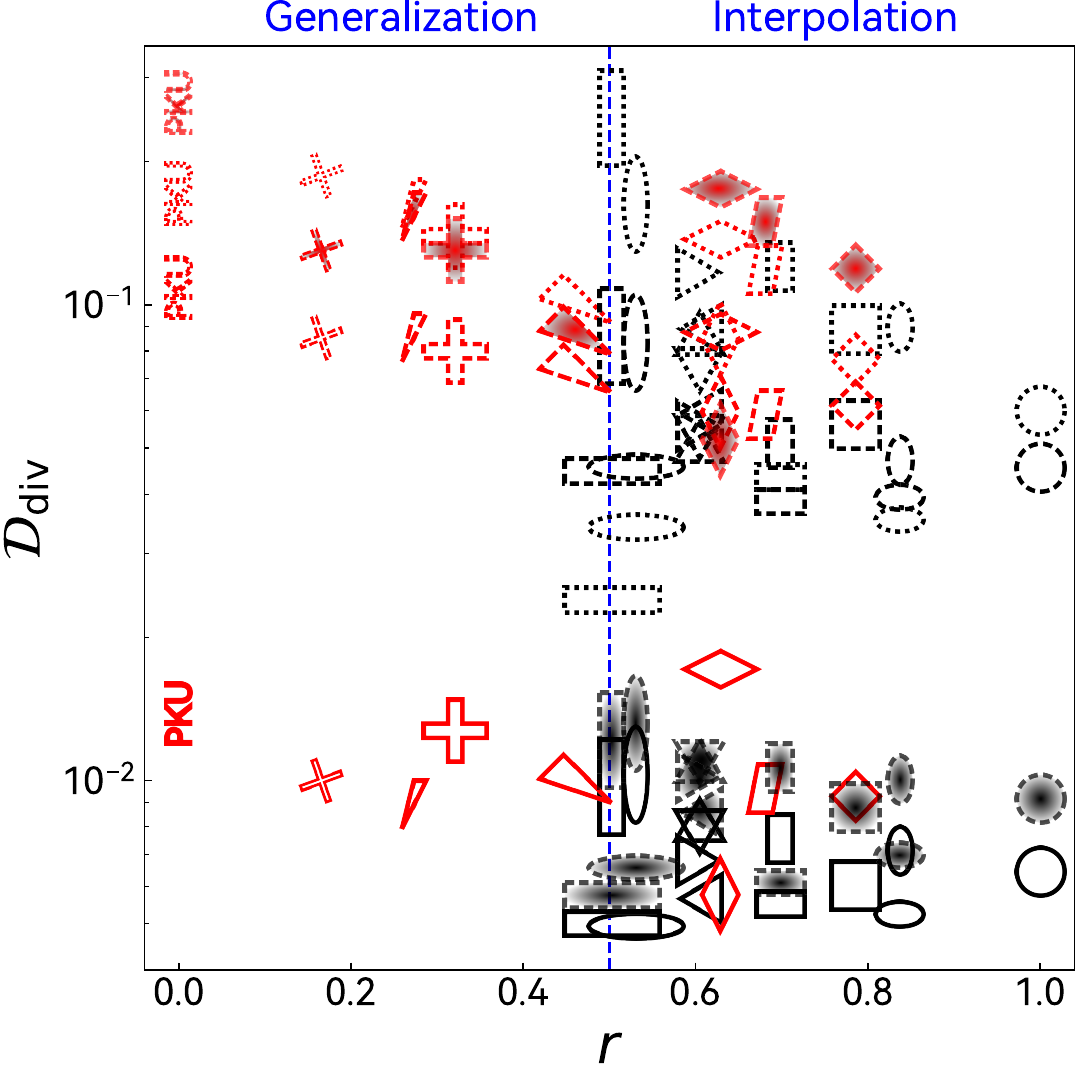}
    \put(5, 95){(b)}
  \end{overpic}
  \caption{
    Comparison of results for different obstacle geometries: 
    (a) normalized $L_1$ error and (b) divergence.
  }
\label{fig:all}
\end{figure}

All geometries' results are summarized in Fig.~\ref{fig:all}. 
The black and red symbols represent the geometries in the training and test sets, respectively. 
The geometries align along the horizontal axis according to their roundness $r$. 
Note that all geometries in the training set have $r \geq 0.5$, while several geometries in the test set have $r < 0.5$, indicating out-of-distribution cases. 

We compare the accuracy of the three ML models in Fig.~\ref{fig:all}(a) using the $L_1$ error of $\lrs{u, v}$ normalized by the difference between the maximum and minimum values in GT, $ L_1 ( [u, v] ) / ( [u, v]_{\max}^{\mr{GT}} - [u, v]_{\min}^{\mr{GT}} ) $. 
The pressure and vorticity are not included due to differences in the order of magnitude. 
The results of the G2F diffusion model are generally better than those of the CNN-a, particularly for the test set.
We also observe that the CNN-a performs well for some slender obstacles, as these cases exhibit weaker vortex shedding. 
Consequently, the wake flows in these instances are close to the time-averaged results by the CNN-a model. 
The effects of the CNN-s model in the training and test sets differ significantly. 
Due to its simpler training strategy, the CNN-s model performs best in the training set but much worse in the test set, indicating a lack of generalization ability. 
The VAE model's score aligns with CNN-a.

We further evaluate the model's realizability using the divergence $\mc{D}_{\mr{div}} = \lrN{\partial_x u + \partial_y v}$ averaged over the domain, which quantifies the discrepancy from the incompressible flow. 
Results in Fig.~\ref{fig:all}(b) demonstrate that $\mathcal{D}_{\mr{div}}$ increases overall as $r$ decreases. 
In all cases, the G2F diffusion model yields a smaller $\mathcal{D}_{\mr{div}}$ than the CNN-based and VAE models, and the rate of increase of $\mathcal{D}_{\mr{div}}$ with decreasing $r$ is smaller. 
The CNN-s model performs badly even in the test set of interpolation cases, such as the diamond and parallelogram. 
The VAE model neglects the divergence-free constraint, too, resulting in physically unrealistic solutions.
This finding confirms the robustness of the G2F diffusion model across various geometries. 
Additionally, flows with $\Rey$ closer to the training-set average (around 200) yield better results, suggesting that data across a broader range of $\Rey$ should be incorporated into the training set to improve model performance.

\section{Conclusion}\label{sec:conclusion}

We propose a G2F diffusion model for predicting flows around obstacles with various geometries. 
The model employs a U-Net architecture with a cross-attention mechanism to incorporate geometry information as a prompt, guiding the reverse diffusion process to generate flow fields. 
We establish a dataset of 2D flows with different geometries for training and testing. The G2F diffusion model is trained using flow fields from simple obstacle geometries and evaluated on both interpolation and generalization tasks.

We assess the G2F diffusion model using simple and complex obstacle geometries present and absent in the training set, respectively. The model results are compared with those from the CNN-based models, VAE, and GT. 
The results demonstrate that the diffusion model outperforms the CNN-based and VAE models, particularly in generating instantaneous flow fields and handling out-of-distribution geometries. 
In particular, the diffusion model successfully captures essential flow features while maintaining physical consistency, even in challenging scenarios.
It reproduces instantaneous flow fields by learning data distribution, while the CNN-a model yields time-averaged data, the CNN-s model struggles with unseen geometries.
Furthermore, the diffusion model can generate more physical results, which is evident from its consistent performance across various geometries, as quantified by the divergence. 

Although the G2F model has a higher computational cost compared to other ML models, its superior physical fidelity justifies this trade-off. 
Our analysis shows G2F excels in capturing vortex dynamics, whereas other models often fall short. 
Despite a 10-second inference time, our G2F model still provides computational savings compared to numerical simulations, which typically require hours of computing time for comparable flow fields. 
This represents a speedup of several orders of magnitude. 
Additionally, the diffusion framework's probabilistic nature allows it to generate diverse, physically plausible flow variations from a single geometric prompt, a capability deterministic models lack.

Beyond the current preliminary study, several issues deserve further investigation. 
In addition to obstacle geometry, $\Rey$ is another key parameter, emphasizing the importance of diverse training data for improving the model's generalization capabilities. 
Future research will expand the dataset to cover a wider range of Reynolds numbers and geometries, exploit the scaling law, and enhance the model's ability to handle diverse flow conditions.
Note that applying our current approach to turbulent flows would require integrating mechanisms to capture the inherent multi-scale nature, for example, through more sophisticated conditioning strategies or hybridizing with turbulence models.

On the methodology perspective, condition implementations should be explored for high-fidelity generation. 
A preliminary test is presented in Appendix~\ref{appendix:phase}.
Furthermore, the ability to generate sequential flow fields is important for practical CFD applications. 
Future work should adapt video generation techniques, built on image diffusion models, to produce temporally coherent sequential flow fields and explore conditional generation across multiple time steps, guided by both geometry and temporal data.

\begin{acknowledgments}

Numerical simulations were performed on the TH-2A supercomputer in Guangzhou, China. This work has been supported in part by the National Natural Science Foundation of China (Nos.~12432010, 52306126, 12525201, and 92270203) and the Xplore Prize.


\end{acknowledgments}

\begin{appendix}
\section{Geometry-and-phase-to-flow diffusion model}\label{appendix:phase}

It is of interest to generate the evolution of the flow field.
One preliminary step is to obtain a snapshot of the flow field at a specified time. 
Consequently, we develop a geometry-and-phase-to-flow (GP2F) diffusion model to generate the flow field from the obstacle geometry and phase information. 
We apply the proper orthogonal decomposition (POD) method~\cite{Berkooz1993} to define the phase $\gamma$ by the weights of the first two POD modes for each case in the dataset.
Here, the phase indicates the normalized time which describes the relative evolution progress in one period and standardizes the time stamps for different geometry cases. 

\begin{figure}[ht]
  \centerline{\includegraphics[width=\textwidth]{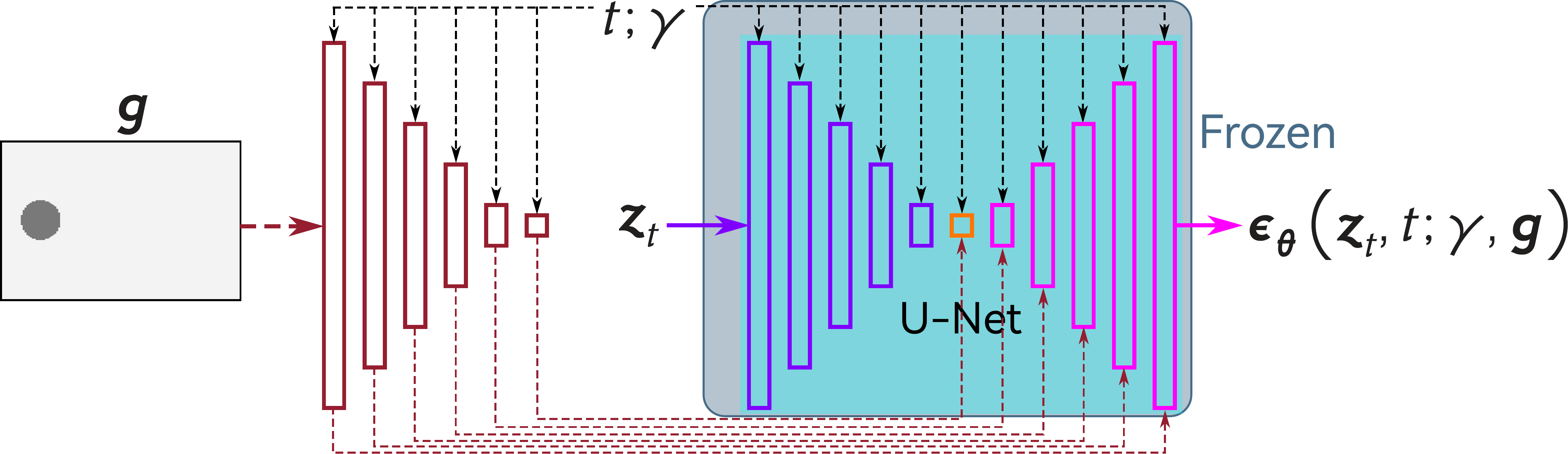}}
  \caption{
  Schematic for the architecture of the GP2F diffusion model. 
  The phase prompt $\gamma$ is injected at every layer of the U-Net. 
  The geometry prompt $\bs{g}$ is injected into the pre-trained U-Net via the same attention mechanism of the G2F model, as shown in Fig.~\ref{fig:DDPM}. 
  During the training of geometry attention blocks, the parameters in U-Net and phase attention blocks are frozen.
  }
\label{fig:UNet-frozen}
\end{figure}

We employ a two-stage training strategy for the GP2F model.  
First, we train a diffusion model conditioned on the phase $\gamma$ using flow around a cylinder, where the $\gamma$ prompt is injected at every layer of the U-Net. 
The U-Net is indicated by the frozen part in Fig.~\ref{fig:UNet-frozen}.  
Next, we encode the geometry prompt $\bs{g}$ through the same attention mechanisms as in the G2F model.
While training the geometry prompt blocks on all the data, we freeze the parameters of the U-Net and phase embedding blocks~\cite{Zhang2023}.

\begin{figure}[ht]
  \centering
  \includegraphics[width=\textwidth]{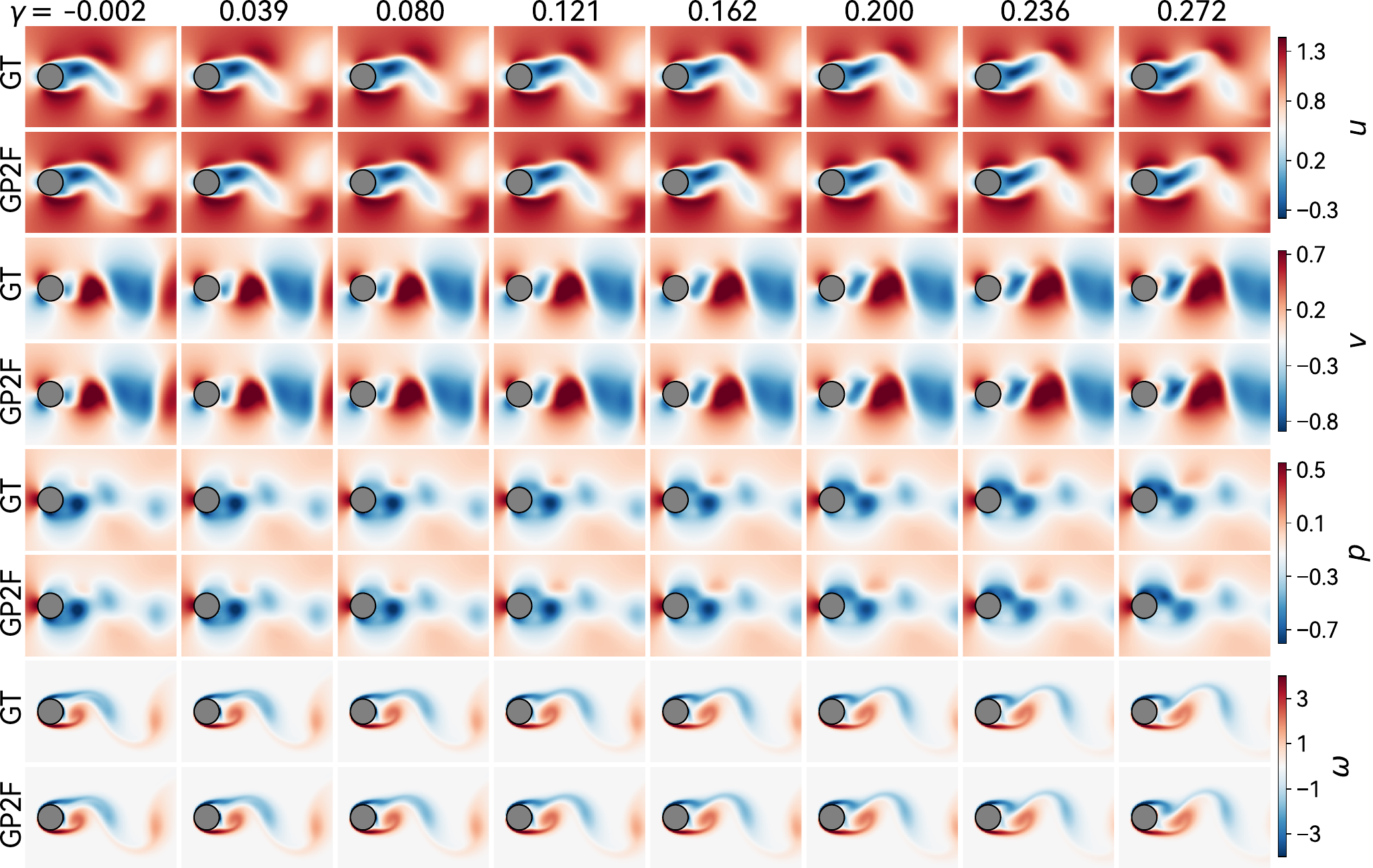}
  \caption{
  The flow fields around a circle generated by the GP2F model over a quarter of the period, compared with the GT on contours of velocities $u$, $v$, pressure $p$, and vorticity $\omega$.
  }
\label{fig:GP2F-0}
\end{figure}

On the cylinder case in the training set, the GP2F model can generate the flow field at a given phase, as shown in Fig.~\ref{fig:GP2F-0}, where eight sequential frames cover a quarter of the period.
Comparison with GT confirms the accuracy of the GP2F model. 
Figure~\ref{fig:GP2F-0}(b) displays the projections of the GP2F results onto the POD modes.
Each point represents a generation result at a given phase.
Points located near the unit circle indicate high-quality generation, demonstrating the high phase accuracy of the GP2F model.

\begin{figure}[ht]
  \centering
  \begin{overpic}[width=0.6\textwidth]{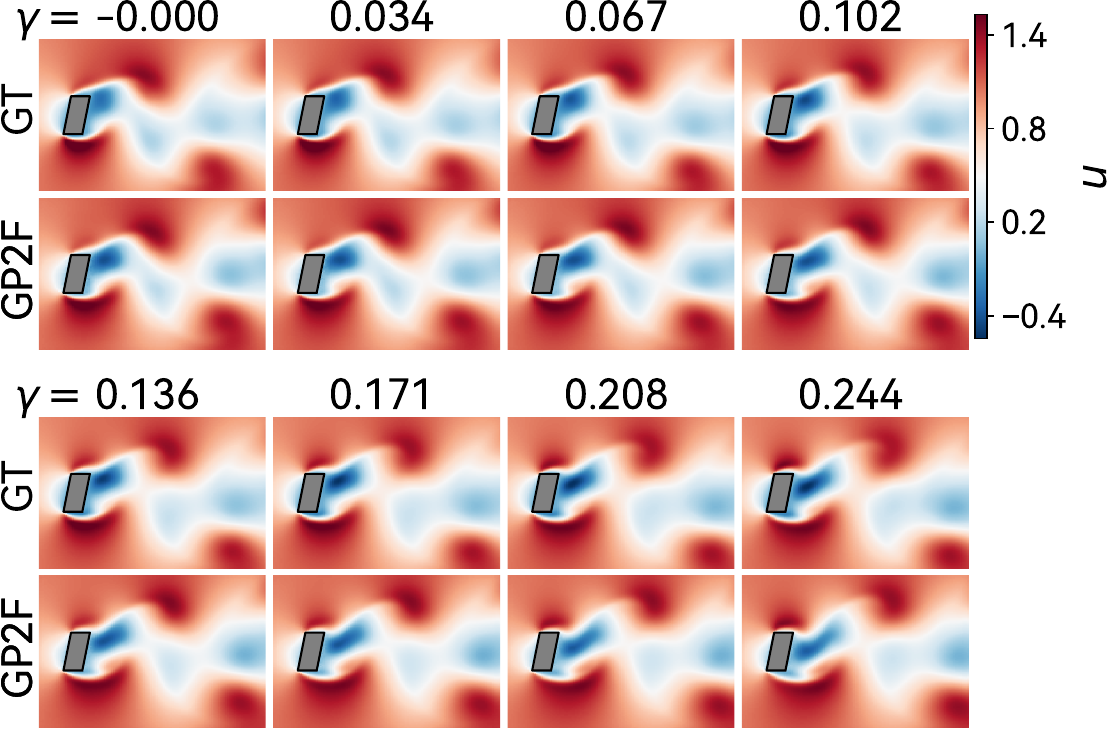}
    \put(-3.5, 62){(a)}
  \end{overpic}\hspace{5pt}
  \begin{overpic}[width=0.36\textwidth]{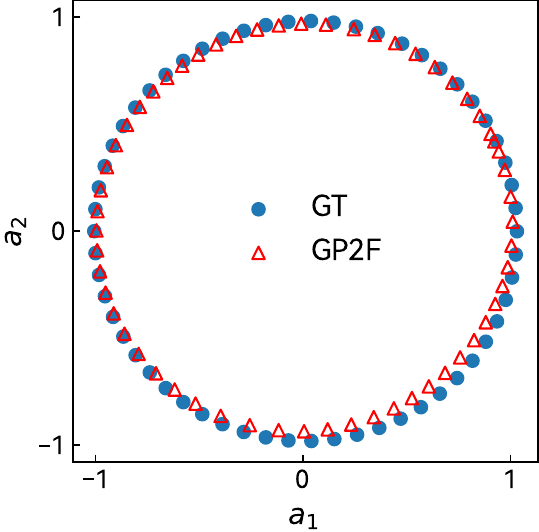}
    \put(0, 93){(b)}
  \end{overpic}
  \caption{
  The flow fields around a parallelogram generated by the GP2F model over a quarter of the period, compared with the GT on  
  (a) contours of velocity $u$; 
  (b) projections onto the POD modes.
  }
\label{fig:GP2F-18}
\end{figure}

The GP2F model is also enable to generate high quality flow field for unseen geometry, like the G2F model.
Figure~\ref{fig:GP2F-18}(a) compares the generated flow around a parallelogram against GT over a quarter of the period.
According to the projections on the POD modes in Fig.~\ref{fig:GP2F-18}(b), the quality of the single-frame flow field is good. 

\begin{figure}[ht]
  \centering
  \begin{overpic}[width=0.6\textwidth]{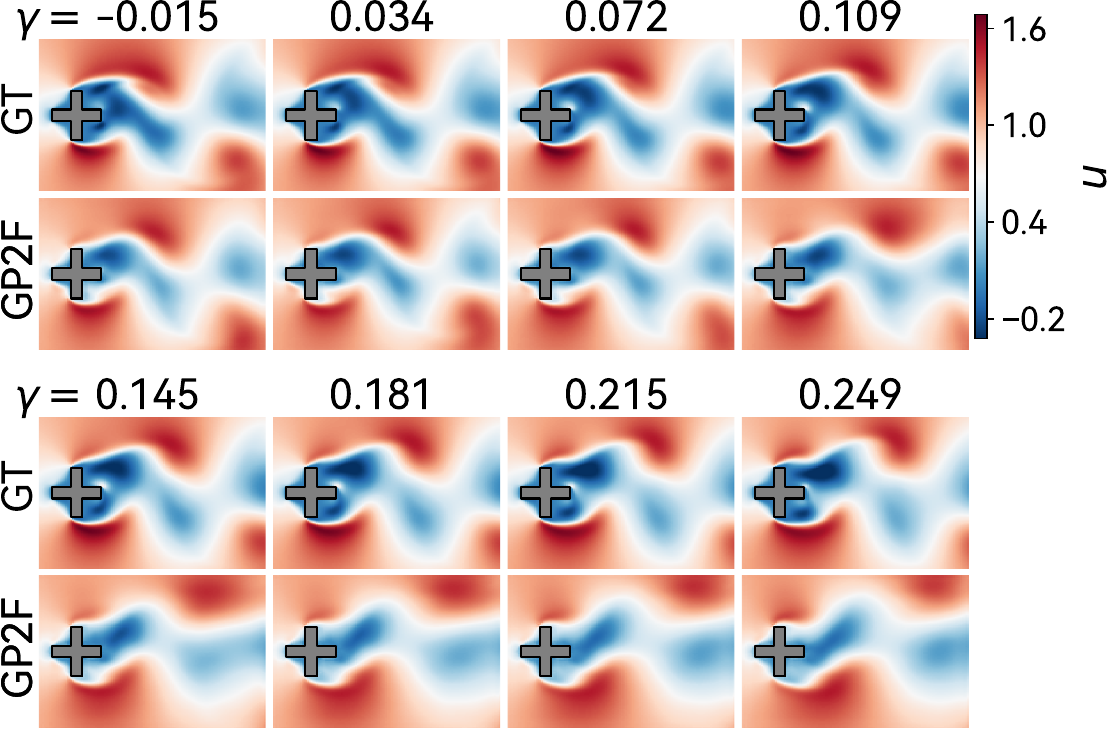}
    \put(-3.5, 62){(a)}
  \end{overpic}\hspace{5pt}
  \begin{overpic}[width=0.36\textwidth]{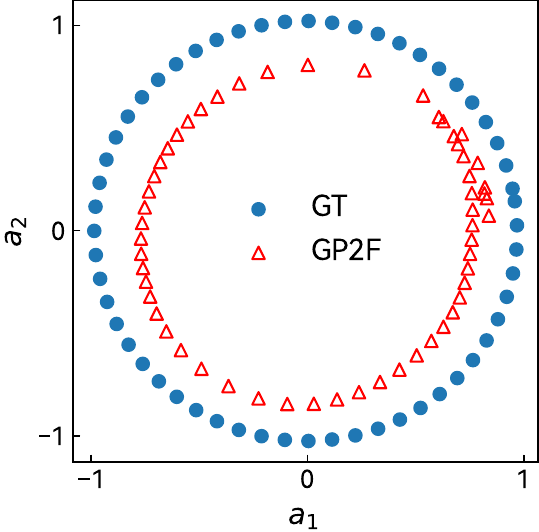}
    \put(0, 93){(b)}
  \end{overpic}
  \caption{
  The flow fields around a cross-shaped obstacle generated by the GP2F model over a quarter of the period, compared with GT on 
  (a) contours of velocity $u$; 
  (b) projections onto the POD modes. 
  }
\label{fig:GP2F-14}
\end{figure}

Figure~\ref{fig:GP2F-14} presents the cross-shaped obstacle case to evaluate the GP2F model on generalization. 
Overall, the GP2F model generates acceptable flow field sequences. 
We also notice the overly smooth flow field and an incorrect recirculation zone near the rear branch, leading to the discrepancies shown in Fig.~\ref{fig:GP2F-14}(b). 
This may be due to the lack of similar geometries in the training set.

\begin{figure}[ht]
  \centering
  \includegraphics[width=\textwidth]{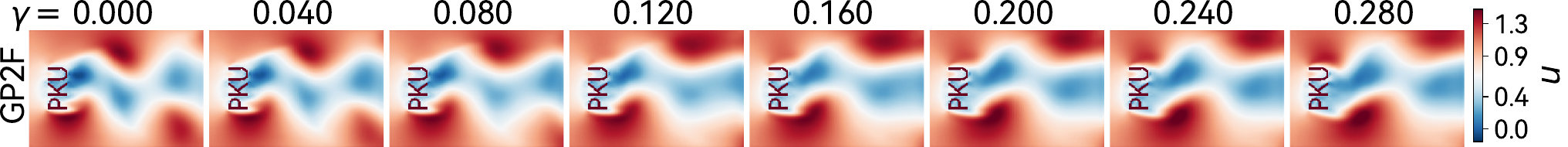}
  \caption{
  Contour of $u$ for the flow field around characters `\textsf{PKU}' generated by the GP2F model over a quarter of the period.
  }
\label{fig:GP2F-19}
\end{figure}

Figure~\ref{fig:GP2F-19} presents the generated flow field sequence over the characters `\textsf{PKU}' by GP2F, illustrating a prospect for further exploration. 
Although the results are not yet precise, we can obtain a preliminary prediction at a low cost. 
Furthermore, the flow field generated by the GP2F model may serve as the initial condition in direct numerical simulations.

Our GP2F model demonstrates that geometry and phase information can effectively determine flow fields across various geometries. We chose diffusion models specifically because their progressive denoising process is particularly valuable for handling the complex, non-linear relations between geometry and resulting flow patterns. The approach to geometry conditioning lays essential groundwork for more complex applications, including the sequential generation.

\section{Physics-informed diffusion model}\label{sec:physics-inform}

To incorporate the physical information in the geometry-to-flow (G2F) diffusion model, we include the losses on the divergence-free condition 
\EQ
    \mc{L}_{\mr{div}} = \lrN{ \pdv{u}{x} + \pdv{v}{y} }^2
\EN
and the vorticity definition 
\EQ
    \mc{L}_\omega = \lrN{ \pdv{u}{y} - \pdv{v}{x} + \omega }^2.
\EN
We combine the basic loss $\mathcal{L}_0$ and physical losses as
\begin{equation}
    \mathcal{L} = \mathcal{L}_{0} + \lambda \lrr{ \mathcal{L}_{\mathrm{div}} + \mathcal{L}_{\omega} },
\label{eq:phy-loss}
\end{equation}
where $\lambda$ is a weighting factor.

\begin{figure}[!htbp]
  \centering
  \includegraphics[width=\textwidth]{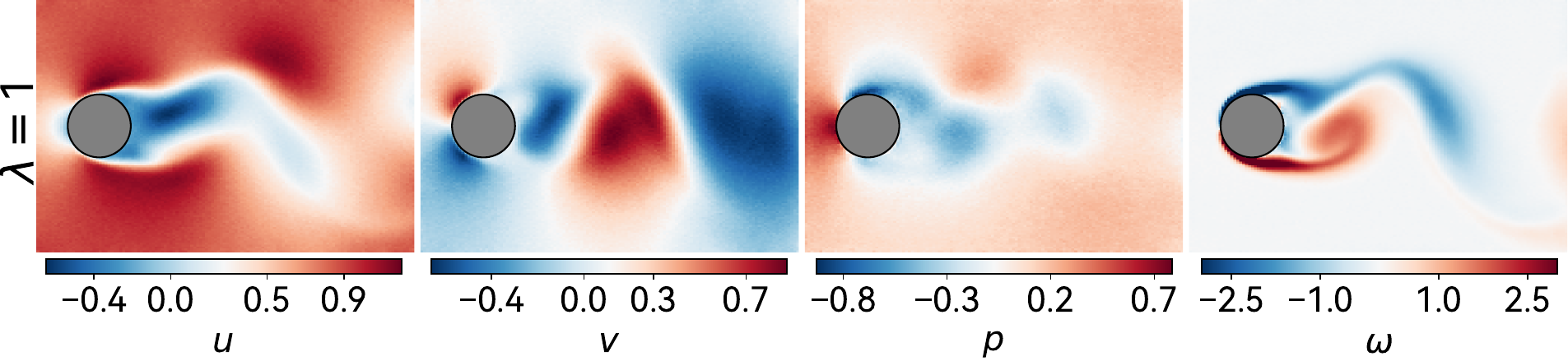}
  \caption{
    Flow fields generated by the physics-informed G2F models trained with physical information.
  }
\label{fig:flow-lambda}
\end{figure}

The loss function $\mc{L}_0$ aims to maximize the likelihood.
However, the additional constraints, $\mc{L}_{\mr{div}}$ and $\mc{L}_{\omega}$, disrupt this objective, leading to a worse result at larger $\lambda$.
Note that Shan \etal~\cite{Shan2024} reported a small weighting factor on the physical constraint in the diffusion model for super-resolution tasks.
We tested with $\lambda = 1$ in \eqref{eq:phy-loss}, on the flow past a cylinder case shown in Fig.~\ref{fig:flow-lambda}.
The physics-informed model performs similarly to the GT and the non-physics-based G2F model.

\begin{figure}[!ht]
  \centering
  \includegraphics[width=0.6\textwidth]{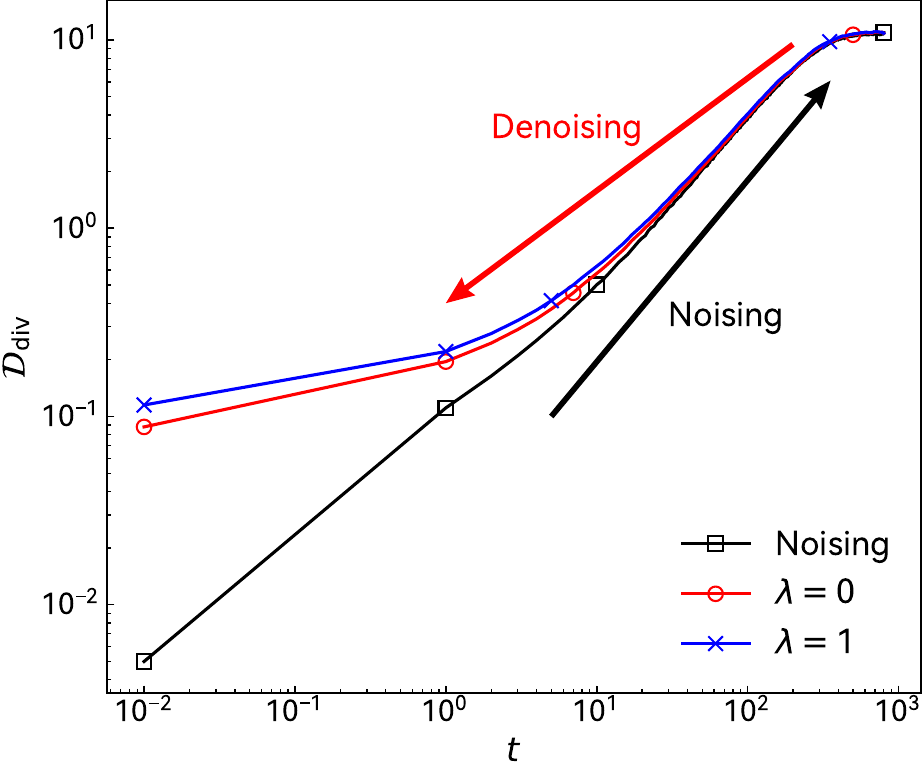}
  \caption{
    Variations of $\mc{D}_{\mr{div}}$ during the forward and reverse diffusion processes, with $\lambda=0, 1$, and $5$. 
    The point at $t = 0$ is replaced with $t = 0.01$ to accommodate the logarithmic scale axis.
  }
\label{fig:div-lambda}
\end{figure}

Figure~\ref{fig:div-lambda} shows the variation of the divergence $\mc{D}_{\mr{div}}$ during the forward and reverse diffusion processes. 
The GT flow fields conform to the divergence-free condition with minimal numerical error. 
As noise is added, $\mc{D}_{\mr{div}}$ increases with step $t$, eventually reaching a maximum value when the field approximates the Gaussian noise.
The reverse diffusion generates a flow field from the Gaussian noise. 
When this field resembles the GT, it naturally conforms to the physical constraints, even without explicit physical information in training.

The $\mc{D}_{\mr{div}}$ variation in reverse diffusion differs from that in forward diffusion. 
Figure~\ref{fig:div-lambda} shows that $\mc{D}_{\mr{div}}$ drops at the same speed as the GT in the first 700 steps during reverse diffusion, then decreases more slowly. 
This behaviour indicates that the diffusion model captures most structures in the early steps, suggesting that completing all diffusion steps may be unnecessary. 
This insight can be leveraged to improve computational efficiency, as noted by Nichol \etal~\cite{Nichol2021}.
To demonstrate the effect of the physical constraints in reverse diffusion, we present the $\mc{D}_{\mr{div}}$ variations with $\lambda = 0$ and 1. 
The final values of $\mc{D}_{\mr{div}}$ do not differ significantly among different $\lambda$ values.

\end{appendix}

\bibliographystyle{elsarticle-num} 
\bibliography{diffCFD}
\end{document}